%% file: zigbeeKUtechrep.tex
\documentclass[a4paper,10pt]{article}
\usepackage{graphicx}
\usepackage{verbatim}
\usepackage{cite}
\usepackage{xspace}
\usepackage{slashbox}
\usepackage{a4wide}
\usepackage{color}
\usepackage{multirow}
\usepackage{wrapfig}
\usepackage[auth-lg,affil-sl]{authblk}

\begin{document}
\input{title}
\input{colophon}

\title{\bfseries Optimizing ZigBee Security\\ using Stochastic Model Checking}
\author[1]{Ender Y\"uksel}
\author[1]{Hanne Riis Nielson}
\author[1]{Flemming Nielson}
\affil[1]{Informatics and Mathematical Modelling, Technical University of Denmark, Richard Petersens Plads, Building 321, DK-2800 Kongens Lyngby, Denmark}
\author[2]{Matthias Fruth}
\author[2]{Marta Kwiatkowska}
\affil[2]{Oxford University Computing Laboratory, Wolfson Building, Parks Road, Oxford, OX1 3QD, United Kingdom}

\maketitle

\begin{abstract}
ZigBee is a fairly new but promising wireless sensor network standard that offers the advantages of simple and low resource communication.
Nevertheless, security is of great concern to ZigBee, and enhancements are prescribed in the latest ZigBee specification: \emph{ZigBee-2007}.
In this technical report, we identify an important gap in the specification on key updates, and present a methodology for determining optimal key update policies and security parameters.
We exploit the stochastic model checking approach using the probabilistic model checker PRISM, and assess the security needs for realistic application scenarios.
\end{abstract}

\section{Introduction}
\label{sec:intro}

Low-cost and low-power wireless sensor networks have a huge potential in many application areas such as home automation, smart energy, industrial applications, etc. 
As an emerging standard, \emph{ZigBee} addresses a significant need in those application areas and besides it is an attractive core technology for ubiquitous computing. 
ZigBee is a \emph{low-rate} standard in terms of cost, power consumption, range, and bandwidth. 
The latest version of the ZigBee specification, that we study throughout this paper is referred to as  \emph{ZigBee-2007} \cite{ZigBee:2007}.

The mere exception in the \emph{low-rate} nature of ZigBee is intended to be in \emph{security}, where implementing security guarantees in an optimal way is of paramount importance. 
However, it is a challenging task to provide secure networking in such a low-rate environment with very limited resources. 

ZigBee-2007 specifies a suite of security services that includes methods for key establishment, key transport, etc.
Although each revision and supporting specifications (such as stack and application profiles) roll out improvements, \emph{key update} strategies and proper determination of related security parameters still remain as gaps in the standard \cite{Yuksel:Nielson:2008}. 

Naturally, we want to ensure that the risk of using a compromised security key in a ZigBee network is as small as possible -- and this calls for updating the key fairly often. 
On the other hand, this operation is computationally expensive and we would not like to perform it too often.
Unfortunately, the ZigBee specification does not give any advice on this (i.e. how and when the key shall be updated) -- it merely states that the security key shall be updated (see Appendix \ref{app:a} for details). 

Given the usual resource limitations in the ZigBee networks, absolute security is often less important than quantifiable trade-offs between security and performance. 
As security is a qualitative concept, realistic analyses require results that are valid with respect to the full behaviour of the systems considered.

\emph{Stochastic model checking} is an automatic technique for verification and performance analysis of stochastic systems. 
Given a stochastic model of a protocol, expressed as a continuous-time Markov chain, it can be used to verify qualitative statements such as ``the maximum probability that the key gets comprised within one year is $0.01$'' and to compute quantitative properties such as ``the expected time to recover from a compromised key''. 
It is also possible to determine optimal solutions to questions such as ``the maximum network size such that confidentiality is maintained with a probability of $0.999$''. 

Applications of quantitative analysis in security domains are rare, as it is usually difficult to quantify or even simulate the underlying problems. 
However, the use of stochastic model checking in \emph{key update} is not only novel, but also greatly effective in that it allows deeper insights than qualitative techniques can provide.
One application of this method is equivalent to a sufficient number of simulation runs, as it covers the full behaviour of the model and delivers provably correct results.

For our experiments, we use the well-established stochastic model checker PRISM \cite{Hinton:Kwiatkowska,Prism:web}, which has been used successfully in over 100 case studies \cite{Kwiatkowska:Norman:2005,Kwiatkowska:Norman:2006} in a wide range of areas such as quality of service, randomised distributed algorithms, security protocols for anonymity, and also contention resolution in wireless network protocols. 

In this paper, we identify gaps left in the specification of the ZigBee-2007 key update protocol and present a methodology for determining optimal key update policies and security parameters with respect to given design constraints (such as confidentiality, recovery time, or update efficiency).
We develop generic models for this protocol and assess individual security needs for specific realistic application scenarios.
For the first time, stochastic model checking is employed for reasoning about ZigBee security services, which allows deeper insights than qualitative techniques or simulation alone. 

Using our beneficial modelling and analysis approach we ask questions and get quantitative answers. 
We focus on optimising key confidentiality, optimising key recovery (from a compromise case) time, and optimising the efficiency of the key updates. 
This paper highlights the main results for all six present application profiles of the ZigBee specification.

This paper is divided into seven sections. 
We present the stochastic models including the key update strategies and application scenarios in Section \ref{sec:models}. 
We investigate the optimisations in key confidentiality, recovery, and efficiency in Sections \ref{sec:confidentiality}, \ref{sec:recovery}, and \ref{sec:efficiency}, respectively. We discuss how to make use of these optimisations in Section \ref{sec:advice} where we give examples based on the experiments we performed.
We present the related work in Section \ref{sec:related}, which is followed by the conclusion. We present the details on the gap in the specification in Appendix \ref{app:a}, and all the probabilistic model checking results for six application profiles, and three key update strategies in Appendix \ref{app:b}.
\section{Stochastic Models for ZigBee}
\label{sec:models}
In this section, we describe how the relevant issues of the ZigBee key update protocol are modelled and analysed. 
This is the basis for overall objective of investigating the relation between different configurable aspects and key update strategies -- with regard to their impact on security -- and  deriving conclusions for the design of security policies and implementations. 
In the remainder of this section, we describe the construction of a generic model for the relevant aspects of the protocol, introduce different key update strategies, and define realistic application scenarios. 
In the next sections, we describe the tool-based optimisation of configurable parameters of the protocol for different application scenarios.

The standard leaves a certain degree of freedom for the design of key update strategies. 
We consider three update strategies for the network key, that we name as: \emph{time-based}, \emph{leave-based}, and \emph{join-based}. 
Our aim for the construction of the model was to make it as simple, modular, and generic as possible, yet detailed enough to allow a realistic analysis and optimisation of four security measures: 
\emph{key confidentiality}, \emph{recovery time}, \emph{network size}, and \emph{update efficiency}. 
\subsection{The Network Model}
First, we present the key points that are necessary for a clear understanding of the development, and we omit all the details which are irrelevant to this study. 
A more detailed discussion of ZigBee security can be found in \cite{Yuksel:Nielson:2008} and surely the ultimate source is the ZigBee documentation \cite{ZigBee:2007, ZigBee:Stack, ZigBeePRO:Stack, ZigBee:HA, ZigBee:SE} which amounts to hundreds of pages including references to several other standards.

ZigBee uses symmetric encryption, the \emph{Advanced Encryption Standard} (AES-128) \cite{AES}, therefore all the security keys are symmetric keys and 128 bits in length. A \emph{Network Key} (\textbf{NK}) is the mere mandatory key in a ZigBee network, which is shared amongst all the devices and used to secure broadcast communications. A \emph{Trust Center} (\textbf{TC}), creates and distributes the NKs. TC is an application running on a ZigBee device, that is unique in every ZigBee network. As a key component of ZigBee security, the TC is assumed to run on a more powerful device (\emph{e.g.} a coordinator) rather than a regular ZigBee end device. Two more types of security keys may exist in a ZigBee network depending on the security configuration: \emph{Master Key} and \emph{Link Key}. Unlike NK, those keys are pairwise shared. In this study we focus on NK as the key type and refer to it as the \emph{key}, and we assume that the devices in the network already acquired the \emph{key}.

The details of the \emph{NK update} protocol is given in the specification and explained in \cite{Yuksel:Nielson:2008}. We assume that when TC updates the key, all the devices in the network successfully update their keys. Compromise of a NK affects all the devices in a network. TC is fully responsible of creating and distributing the NK, therefore there is no role of the devices in NK establishment. 

Our model of the ZigBee network as a continuous-time Markov chain (\textbf{CTMC}) in PRISM is shown in Table \ref{tab:network}. The states are described by two variables:
\begin{itemize}
\item {\tt Size} holds the number of devices currently in the network, and
\item {\tt Comp} tells whether or not the network key is currently compromised.
\end{itemize}

\begin{table}[b]
\hrule \small
\begin{verbatim}

module NETWORK

  Size: [0..Max] init Max;
  Comp: bool init false;

  [leave]  Size>0   -> R_leave*(1-P_comp)*Size: (Size'=Size-1);
  [leaveC] Size>0   -> R_leave*P_comp*Size:     (Size'=Size-1) & (Comp'=true);
  [join]   Size<Max -> R_join*(Max-Size):       (Size'=Size+1);
  [reset]  true     ->                          (Comp'=false);

endmodule
\end{verbatim}
\hrule
\caption{Network model}\label{tab:network}
\end{table}
The TC will know the value of {\tt Size} but obviously not that of {\tt Comp}. At this point we should mention that we assume the network is deployed and started with its maximum number of devices, {\tt Max}. In other words, we tend to preserve the number of devices in the network. Actually, this assumption originates from real life examples of ZigBee products such as buying a set of sensor devices each of them having a different task for your house, or deploying sensor devices for energy metering to an industrial facility. Naturally, the key is fresh, in other words not compromised in the beginning.
In our model we shall be interested in the following actions:
\begin{itemize}
\item {\tt leave}: A device leaves the network with rate {\tt R\_leave}.
\item {\tt join}: A device joins the network with rate {\tt R\_join}.
\item {\tt reset}: The network key is updated (by all devices in the network).
\end{itemize}
When a device leaves the network it still may hold a valid network key so there is a risk that it is compromised; to model this we introduce a variant of the action {\tt leave}:
\begin{itemize}
\item {\tt leaveC}: A device leaves the network while compromising the key with probability {\tt P\_comp}.
\end{itemize}

The values of the rates {\tt R\_leave} and {\tt R\_join} will be determined by the actual application scenario as will the maximal size {\tt Max} of the network and the probability {\tt P\_comp} of the key being compromised. Throughout the paper we shall assume that a \textbf{time unit} in the CTMC correspond to one {\bf day}.
\subsection{Key Update Strategies}
Our network model in Table \ref{tab:network} is operating in an environment that controls when the network key is updated. We shall consider three strategies for doing so. The ZigBee-2007 specification uses the word ``\emph{periodically}'' when referring to the key update issue but gives no further guidelines. Our three approaches are:

\textbf{Time-based key update.}
The \emph{time-based} models are built on the traditional concept of updating the key after a certain amount of time elapsed. The key is updated after a predefined  \emph{key expire time}. 

\textbf{Leave-based key update.}
The \emph{leave-based} models are considering the leave events in the network. The key is updated when a predefined number of devices has left the network. In practice, when a device leaves the network it may still own a valid key, hence a device leave presents a security risk. To the best of our knowledge, this is a novel key update strategy that we propose in this paper. A counter in the Trust Center keeps the number of the devices left (or removed from) the network. When this number reaches the predefined threshold value, all the keys in the network are updated and the counter is reset to zero. The idea here is to have a key update strategy that is inspired by the nature of the wireless sensor networks where number of exchanged messages can be very low compared to the number of join (or leave) events.
 
\textbf{Join-based key update.}
The \emph{join-based} models are considering the join events in the network. The key is updated when a predefined number of new devices has joined the network. A new device that joins the network presents a security risk since it may become a legitimate attacker. This strategy is also our own proposal, and the idea is very similar to the Leave-based key update. In this case, the counter would keep the number of joining devices, and the threshold would be set as a limit for this value.

For each of these strategies, we have an environment that we present in Table \ref{tab:env}:

\begin{itemize}
\item {\tt ENV\_time}: key reset whenever {\tt T\_time} months (i.e. 30*{\tt T\_time} time units) have passed.
\item {\tt ENV\_join}: key reset whenever {\tt T\_join} devices have joined the network.
\item {\tt ENV\_leave}: key reset whenever {\tt T\_leave} devices have left the network.
\end{itemize}

\begin{table}[!h]
\hrule
\begin{quote}\small
\begin{verbatim}

const int T_time;

module ENV_time

  [leave]  true ->                true;
  [leaveC] true ->                true;
  [join]   true ->                true;
  [reset]  true -> 1/(30*T_time): true;

endmodule

const int T_join;

module ENV_join

  C_join: [0..T_join] init 0;

  [leave]  C_join<T_join ->           true;
  [leaveC] C_join<T_join ->           true;
  [join]   C_join<T_join ->           (C_join'=C_join+1);
  [reset]  C_join=T_join -> R_reset:  (C_join'=0);

endmodule

const int T_leave;

module ENV_leave

  C_leave: [0..T_leave] init 0;

  [leave]  C_leave<T_leave ->          (C_leave'=C_leave+1);
  [leaveC] C_leave<T_leave ->          (C_leave'=C_leave+1);
  [join]   C_leave<T_leave ->          true;
  [reset]  C_leave=T_leave -> R_reset: (C_leave'=0);

endmodule
\end{verbatim}
\end{quote}
\hrule
\caption{Environments}\label{tab:env}
\end{table}

Thus the {\tt Network} module plays the active role for the {\tt leave}/{\tt leaveC} and {\tt join} actions whereas the environment {\tt ENV\_time}/{\tt ENV\_join}/{\tt ENV\_leave} plays the active role for the {\tt reset} action. The values of the thresholds {\tt T\_time}, {\tt T\_join}, and {\tt T\_leave} are parameters to the key update strategies and will be varied below; for simplicity we shall take
\begin{quote}\small
\begin{verbatim}
const double R_reset   = 1/24;   
\end{verbatim}
\end{quote}
reflecting that the key reset in average takes one hour.
\subsection{The Application Scenarios}
\label{sec:appsce}
ZigBee has six different application profiles and up to now only two of them are finalized, \emph{Home Automation} \cite{ZigBee:HA} and \emph{Smart Energy} \cite{ZigBee:SE}. The remaining application profiles that are expected to be finalized and released soon are \emph{Commercial Building Automation}, \emph{Personal, Home and Hospital Care}, \emph{Telecom Applications}, and \emph{Wireless Sensor Applications}. Before presenting the details of the scenarios, we start by giving our assumptions that are common to all scenarios.

As we aim to estimate system dimensions and other parameter values (or value ranges) for different applications, we use application profiles that are defined by ZigBee. Below we explain the settings for the three application profiles that we focus on in this study. The remaining application profiles are also studied in Appendix \ref{app:b}. For each application profile we first present the information gathered from a professional ZigBee expert \cite{Cragie} in a paragraph and then present the parameters that we used in our models with the values that we determined from that information.

\paragraph{The Home Automation Scenario:}
In this scenario, the network is \emph{fairly static} such that it is likely that devices such as light switches and luminaries, once commissioned, would remain in place for a longer period. The network size is in general less than 50 devices. The environment is relatively insecure, and to reflect this we shall say the key is compromised in $1\%$ of the cases. A device may leave the network for reasons such as a break down or flat battery, and most likely, it will be replaced shortly after. We shall assume that each device leaves the network once a year but it will be replaced within a week. Based on these assumptions we specify the remaining constants as follows:
\begin{quote}\small
\begin{verbatim}
const int    Max      = 20;    // maximal size of network
const double R_join   = 1/7;   // average join: 1 every week
const double R_leave  = 1/365; // average leave: 1 per year
const double P_comp   = 1/100; // risk of key compromise
\end{verbatim}
\end{quote}

\paragraph{The Smart Energy Scenario:}
In this scenario, the network is \emph{static}, so the devices rarely leave the network. The network size is in general 3-5 devices, including an Energy Service Portal in a meter, a Programmable Communicating Thermostat, and a display. The environment is highly secure, and to reflect this the risk of key compromise is very low. Once a device has left the network, it will be replaced shortly. Based on these assumptions we specify the remaining constants as follows:
\begin{quote}\small
\begin{verbatim}
const int    Max      = 5;         // maximal size of network
const double R_join   = 1/7;       // average join: 1 every week
const double R_leave  = 1/(5*365); // average leave: 1 per five years
const double P_comp   = 1/100000;  // risk of key compromise
\end{verbatim}
\end{quote}

\paragraph{The Commercial Building Automation Scenario:}
Also in this scenario, the network is \emph{fairly static} but its size is somewhat higher than in the previous scenarios although it is in general less than 200 devices. The environment is relatively secure, and also here left devices will be replaced within a short period. Based on these assumptions we specify the remaining constants as follows:
\begin{quote}\small
\begin{verbatim}
const int    Max      = 100;    // maximal size of network
const double R_join   = 1/7;    // average join: 1 every week
const double R_leave  = 1/365;  // average leave: 1 per year
const double P_comp   = 1/1000; // risk of key compromise
\end{verbatim}
\end{quote}

Obviously, application profiles can be customized easily and the models can also be used for different type of networks.

\section{Optimising Key Confidentiality}
\label{sec:confidentiality}
Using our models and input values for specific ZigBee application profiles, we can obtain the probability of being in a state where the key is compromised at a specific time instant. Thus, we also show that stochastic model checking can be efficiently used in determining the most appropriate key update threshold (e.g. time period) for an intended security level. Below is the continuous stochastic logic (\textbf{CSL}) \cite{CSL} formula that we use for this series of experiments:

\begin{quote}\begin{center} {\tt P=? [ F[30*T,30*T] Comp ] } \end{center}\end{quote}

\begin{center}
\begin{figure}[!h]
    \begin{tabular}{cc}
      \includegraphics[width=8.5cm]{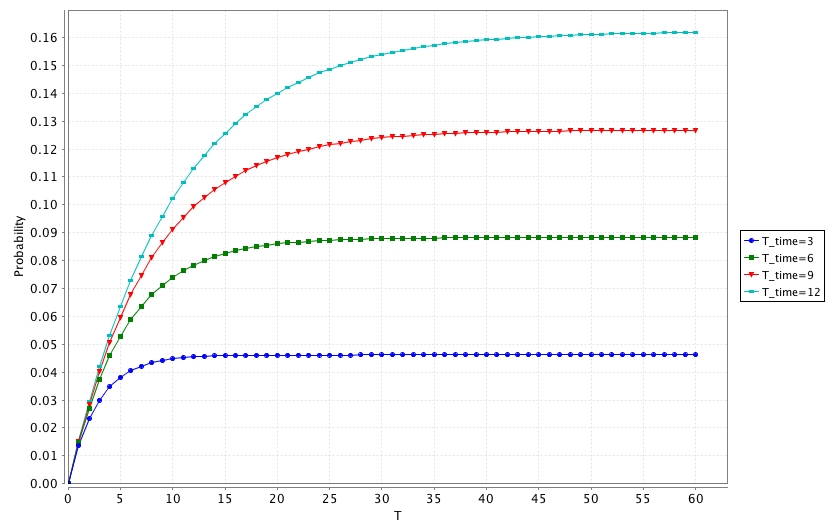} & \includegraphics[width=8.5cm]{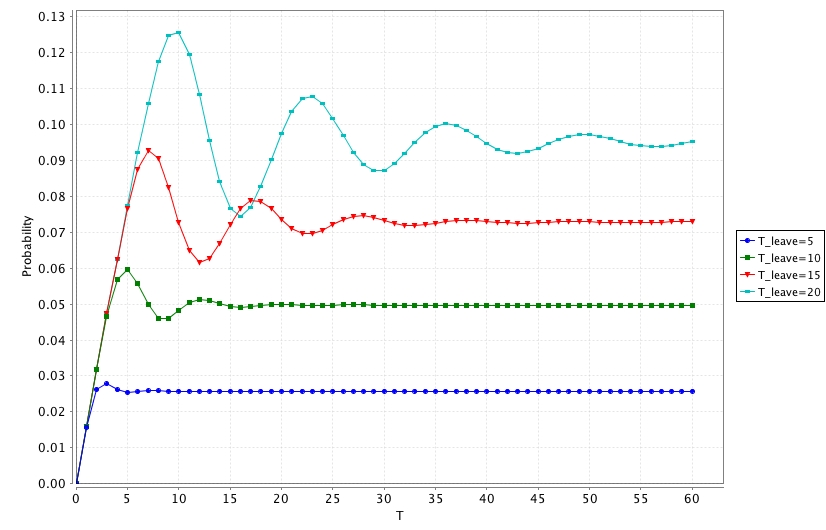}       \\
      \small{a) Time-based key update} & \small{b) Leave-based key update} \\
    \end{tabular}
  \caption{Probability of key compromise in \emph{Home Automation} using a) \emph{time-based} b) \emph{leave-based} key update.}
  \label{fig:ha-q1}
\end{figure}
\end{center}

As an example, we present the results for the \emph{Home Automation} scenario, using the \emph{time-based} and \emph{leave-based} key update strategies in Fig. \ref{fig:ha-q1}. The x-axis shows the month {\tt T} of interest and spans a 5 years period. On the y-axis we have the probability (or risk) that the key is compromised at that time instant.  The curves represent different thresholds (i.e. {\tt T\_time} or {\tt T\_leave}) for resetting the key.

In Fig. \ref{fig:ha-q1}-a, we present the results using \emph{time-based} key update, where we reset the key every 3 months, every 6 months, etc. up to every 12 months. We see that the risk of the key being compromised at the first month instant is about 1,5\% almost independently on the frequency of the key resets -- this is in no way surprising as each of the ({\tt Max}=) 20 devices may leave the network any of the 30 days with a rate of ({\tt R\_leave}=) 1/365 and in ({\tt P\_comp}=) 1/100 of these cases the key will be compromised (and indeed 20*30*1/365*1/100 amounts to 0,016) and at the same time the chance of having had a key reset within the first month is low. In addition, if the key is reset on average once a year (blue curve) then the risk is about 11\% that it will be compromised after a year but it is only 4,5\% if the key is reset once every three months (capri curve).

The results for the same scenario using the \emph{leave-based} key update are presented in Fig. \ref{fig:ha-q1}-b. Here, we have a curve for each of the key update threshold (i.e. {\tt T\_leave}) of 5, 10, 15 and 20 devices leaving the network. Again, the risk of the key being compromised at the first month instance is about 1,6\% almost independently of the threshold. If the key is reset whenever 5 devices have left the network (navy blue curve) then the risk of the key being compromised after 12 months is just 2,5\% which is somewhat better than the time based environment (for that specific initial value set of the parameters). 

We can now say that if the key is reset after every 20 devices have left the network (capri curve) then the risk of the key being compromised after 12 months will be about 10\% -- also this is slightly better that updating the key once a year. However, stability seems only to have been reached after more than 5 years of operation. 

The reason behind the fluctuations before the stabilisation is the nature of \emph{leave-based} key update. Sticking to the example of the black curve, we can see that peak points are approximately where we reach the key update threshold. The threshold value of 20 is reached after almost 9 months which is where the risk is at highest and then the key is reset causing a drop sequence in probability. The fluctuations disappear as time goes by, still having the peaks before the key is reset.

Naturally, we are also interested in the preservation of confidentiality in the \emph{long-run} or \emph{equilibrium}. Using stochastic model checking, we can obtain the steady-state probability of key compromise for different key update thresholds. Below is the CSL formula that we use for this series of experiments:

\begin{quote}\begin{center} {\tt S=? [ Comp ] } \end{center}\end{quote}

\begin{center}
\begin{figure}[!h]
    \begin{tabular}{cc}
      \includegraphics[width=8.5cm]{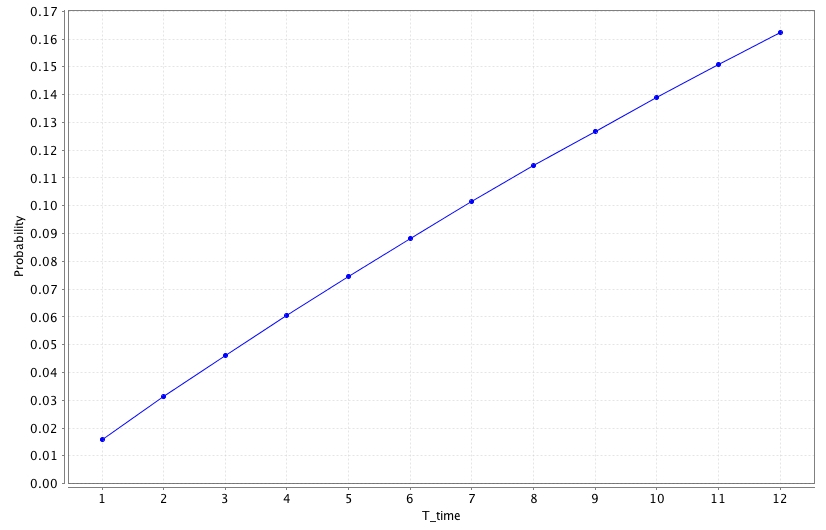} & \includegraphics[width=8.5cm]{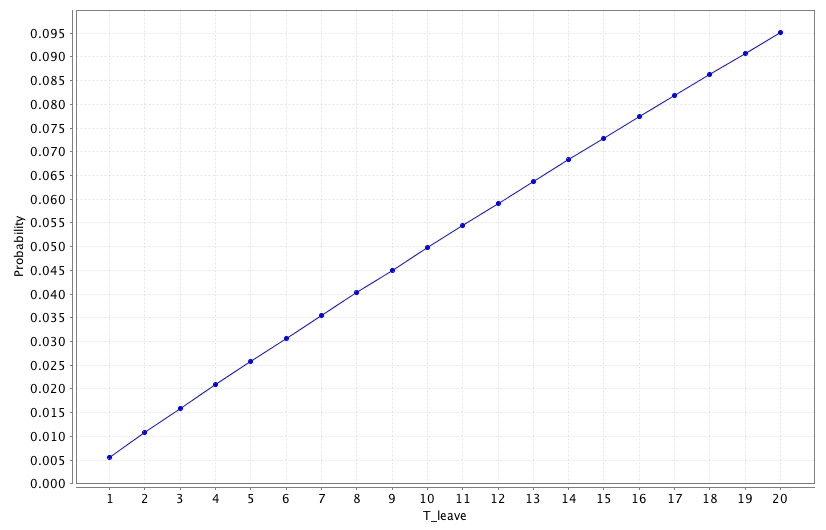}       \\
      \small{a) Time-based key update} & \small{b) Leave-based key update} \\
    \end{tabular}
  \caption{Steady-state probability of key compromise in \emph{Home Automation} using a) \emph{time-based} b) \emph{leave-based} key update.}
 \label{fig:ha-t-q2}
\end{figure}
\end{center}

We present the results for the \emph{Home Automation} scenario, using the \emph{time-based} and \emph{leave-based} key update strategies in Fig. \ref{fig:ha-t-q2}. Now the x-axis shows the threshold values for reset (i.e. the interval {\tt T\_time} in which the reset is performed for the time-based, and the number of left devices {\tt T\_leave} that is required to trigger the reset for the leave-based) and the y-axis shows the steady state probability that the key is compromised.

As an example we see that if the key is reset every three months then the probability of a compromised key is about 4,5\% thereby confirming the the result of Fig. \ref{fig:ha-q1}-a after the 12 year period. If a reset only happens once a year then the steady state probability is 16\% which verifies what we saw in Fig. \ref{fig:ha-q1}-a after the stabilisation.

\section{Optimising Recovery Time}
\label{sec:recovery}
A security key may get compromised for several reasons, and eventually it will be updated. However, the time needed to recover from this needs to be optimized in such a way that the network is not without protection for a long time period. Thus, we are interested in the risk that it takes more than a specific amount of time to recover from a compromise of the key. Below is the CSL formula that we use for this series of experiments:

\begin{quote}\begin{center} {\tt P=? [ Comp U>=(30*T) !Comp \{Comp\}\{max\} ] } \end{center}\end{quote}

\begin{center}
\begin{figure}[!h]
    \begin{tabular}{cc}
      \includegraphics[width=8.5cm]{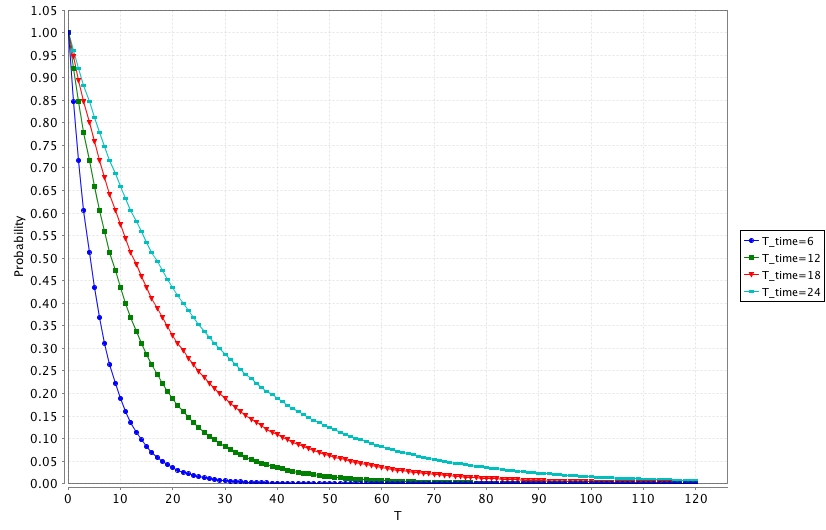} & \includegraphics[width=8.5cm]{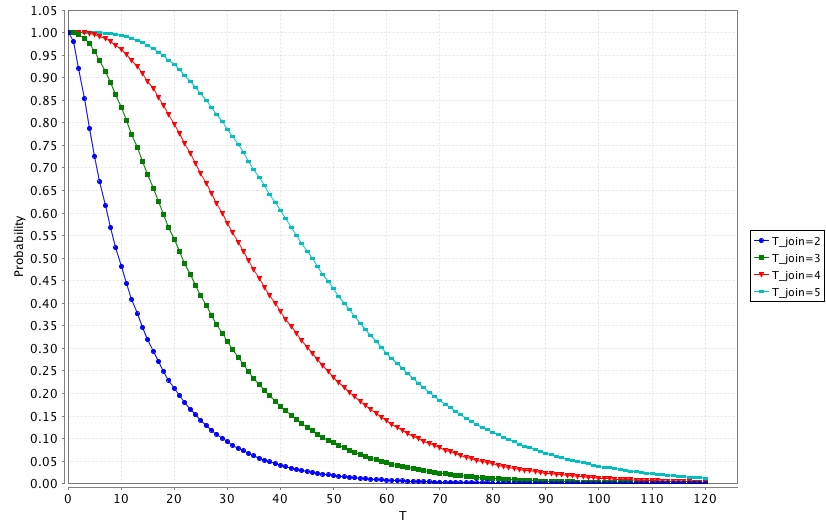}       \\
      \small{a) \emph{Commercial Building Automation}, Time-based} & \small{b) \emph{Smart Energy}, Join-based} \\
    \end{tabular}
  \caption{The probability of key recovery takes more than {\tt T} months.}
  \label{fig:rec-q3}
\end{figure}
\end{center}

We present the results for two different scenarios, using two different key update strategies in Fig. \ref{fig:rec-q3}. In both cases, the x-axis shows the number {\tt T} of months that the key will be compromised before it is reset. On the y-axis we have the probability (or risk) that we have to wait that long. The curves to the left represent different intervals {\tt T\_time} for resetting the key in the \emph{time-based} strategy, whereas the curves to the right represent different thresholds {\tt T\_join} for resetting the key in the \emph{join-based} strategy.

In Fig. \ref{fig:rec-q3}-a we focus on the \emph{Commercial Building Automation} using the \emph{time-based} key update. We observe that if the key is reset on average every 18 months (red curve) then the risk is about 89\% that the key will be compromised for more than 2 months before it is reset whereas it is about 51\% that the key will be compromised for a full year. On the other hand, if the key is reset every 6 months (navy blue curve) then the risk is about 72\% that the key is still compromised after 2 months whereas it is just less than 14\% that it is still compromised after a year. The curves cover a period of 10 years.

The results of the \emph{join-based} key update represent a different pattern than the \emph{time-based} one. In Fig. \ref{fig:rec-q3}-b, we present the results for the \emph{Smart Energy} scenario. In the case where the key is reset after a total of 4 devices joined the network (red curve), we see that the probability of the key being compromised for more than 2 months is almost 100\% -- and with 94\% probability the key will be compromised for at least a full year before being reset. Again the curves cover a period of 10 years.

As a special case, when the key is reset after each join (not shown in the figure), we find out that the risk of a key recovery that takes more than 2 months is negligible.

\section{Optimising Efficiency of Key Updates}
\label{sec:efficiency}
In this section, we investigate the extent to which the key updates triggered by the various threshold values are indeed needed, that is, will the key really be compromised when the resets occur. We shall classify the key updates as \emph{useful} and \emph{useless} resets. A \emph{useful reset} is a key update that is applied after a key gets compromised, therefore recovering the key. Similarly, a \emph{useless reset} is a key update that was not necessary because the key was not compromised. Naturally, we want the percentage of useless resets to be as small as possible because a key update is a costly action for ZigBee devices. We keep the scenarios and the settings of the previous sections, and find out how efficient were our values. To answer this question we shall introduce three transition rewards in our PRISM model as shown in Table \ref{tab:rewards}. Below are the CSL formulae that we use for specifying steady-state reward properties in this series of experiments:

\begin{quote}\begin{center} {\tt (100 * R\{"Useful\_Resets"\}=? [ S ]) / R\{"All\_Resets"\}=? [ S ] } \end{center}\end{quote}
\begin{quote}\begin{center} {\tt (100 * R\{"Useless\_Resets"\}=? [ S ]) / R\{"All\_Resets"\}=? [ S ] } \end{center}\end{quote}

\begin{center}
\begin{figure}[!h]
    \begin{tabular}{cc}
      \includegraphics[width=8.5cm]{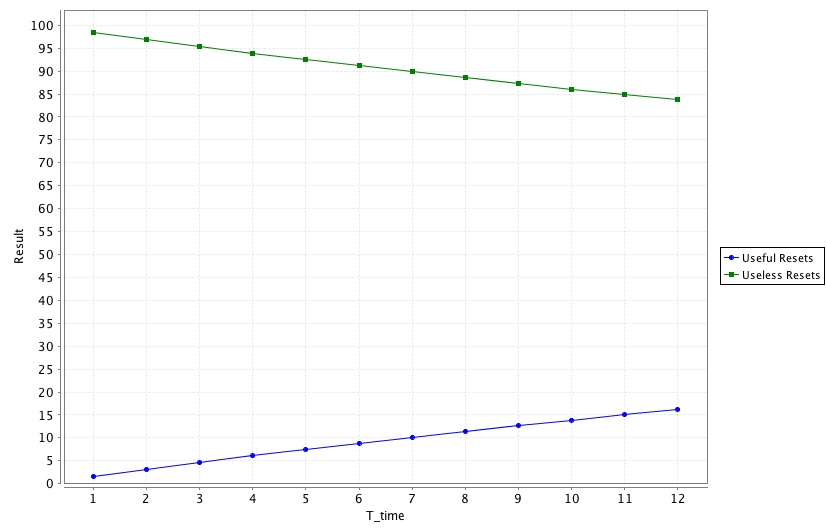} & \includegraphics[width=8.5cm]{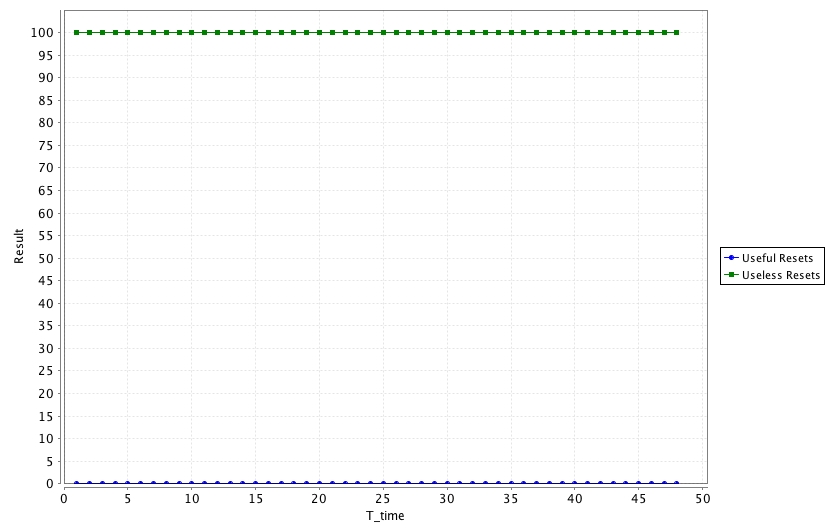}       \\
      \small{a) Home Automation} & \small{b) Smart Energy} \\
    \end{tabular}
  \caption{The efficiency of the key updates using \emph{time-based} key update strategy.}
  \label{fig:eff-q4}
\end{figure}
\end{center}

\begin{table}[!h]
\hrule
\begin{quote}\small
\begin{verbatim}

rewards "All_Resets"
  [reset] true:  1;
endrewards

rewards "Useful_Resets"
  [reset] Comp:  1;
endrewards

rewards "Useless_Resets"
  [reset] !Comp:  1;
endrewards
\end{verbatim}
\end{quote}
\hrule
\caption{Rewards}\label{tab:rewards}
\end{table}
In Fig. \ref{fig:eff-q4}, we present the results of the \emph{time-based} key update strategy for \emph{Home Automation} and \emph{Smart Energy} application profiles. In both cases, the y axis represents the percentage of resets, and the x axis represents the threshold values (i.e. {\tt T\_time}) for key update. Obviously, as the threshold value grows, the number of resets will drop. However, it is not obvious how the \emph{percentage} of the resets will change depending on the threshold values in different key update strategies and scenarios. Fig. \ref{fig:eff-q4}-a shows that the percentage of useless resets drops about 15\% between thresholds of 1 month and 12 months. The change in Fig. \ref{fig:eff-q4}-b is much less visible because of the properties of the selected application profile, even though the observation period is 4 years. Here, we can see that the trade-off between performance and security is harsher in \emph{Home Automation}. The lines in \emph{Smart Energy} are indeed converging with a very slow rate. 

\section{Deriving Advice from Stochastic Model Checking}
\label{sec:advice}
Obviously, it is not trivial to derive conclusion from the stochastic model checking results on the key update regarding confidentiality, recovery, and efficiency. For instance, the more efficient configuration is not the more secure one. To overcome such dilemma, designers should decide on the priorities of the system and select appropriate security parameters. In this section, we give examples on how to decide on the optimum key update strategy and key update threshold. We start by choosing the application scenario, which depends on the type of the sensor devices and the objectives of the network. To choose the application scenario is fairly easy because ZigBee specification already has  specialized application profiles that the designers and developers are supposed to make use of. Then comes the requirements which can be about security, i.e. confidentiality, and performance, i.e. power consumption. Obviously, an extra key update causes an unwanted power consumption and therefore would drain the batteries earlier than expected. After carefully specifying the requirements, we can exploit stochastic model checking on finding answers to our questions. At this point, picking a collection of different key update strategies and a set of threshold values would make everything easier. Model checking results will point us the appropriate threshold values if they exist, and of course different behaviours of different key update strategies. Getting all this information, the rest is evaluating all the solutions for all the requirements and conclude on the solution that satisfies all the requirements. In the rest of this section, we have two examples that shows how we can get advice from stochastic model checking. In addition, it might be seen as a competition between different key update methods and we see how a method can beat another when different environment conditions and requirements exist.

\fbox{\textit{\textbf{Example 1}}:} We start by a simple example, where we can show the power of stochastic model checking in figuring out the best strategy, and also parameter selection for that strategy.

{\bf Step I:} We first determine our application scenario. Since the sensor applications might have different application profiles, it is relatively easy to make this selection. In this example, we assume that the application scenario is the \emph{Personal, Home and Hospital Care} scenario (PHHC). We left the details of this scenario to the appendix (see Appendix \ref{app:sce}) and did not use any results up to now, however the information that we need to know is given below:
\begin{itemize}
\item maximum number of devices: \textit{500 devices.}
\item average join: \emph{1 device per week.}
\item average leave: \emph{1 device per month.}
\item risk of key compromise: \emph{0.01\%.} 
\end{itemize}
{\bf Step II:} Now, we determine our requirements regarding the security parameters. For the sake of simplicity, we only define one requirement in this example and it would be about key compromise.
\begin{itemize}
\item R1 - \emph{the probability that the key is compromised must be lower than \underline{0.1\%}, \underline{at any time}.}
\end{itemize}
{\bf Step III:} Then, we can determine the parameter values that would satisfy the requirements in the previous steps. To do this, all we need is the graphical results of stochastic model checking. For this simple example, all we need to do is to check the long run behaviour of the key update mechanism and we check the results for specific time instants. Notice that all the graphical results that we use in this example are in Appendix \ref{app:answers}, and to keep the example simple we restrict the set of threshold values to be $\{1,2,3,4\}$ for the time-based, $\{5,10,15,20\}$ for the leave-based, and $\{5,10,15,20\}$ for the join-based key update strategy.
\begin{itemize}
  \item \textbf{Time-based: }Looking at the steady-state probabilities (Fig. \ref{tab:q2-t-2}-b) we figure out that even the smallest periodical key update threshold in terms of months, i.e. 1 months, can not satisfy R1. This situation is also visible in the results of transient probabilities (Fig. \ref{tab:q1-t-2}-b). Therefore, \emph{this strategy does not satisfy the requirement}.
  \item \textbf{Leave-based: }Looking at the steady-state probabilities (Fig. \ref{tab:q2-l-2}-b) we figure out that in order to satisfy R1 the threshold value should be less than or equal to 10 (devices). Then, checking the transient probabilities we find out that this situation is consistent and for the set of thresholds we have in Fig. \ref{tab:q1-l-2}-b), the values 5 and 10 will perfectly satisfy R1.
  \item \textbf{Join-based: } If we only checked the steady-state probabilities (Fig. \ref{tab:q2-j-2}-b), then we would conclude that any threshold values less than or equal to 10 would satisfy R1. However, when we check the transient probabilities (Fig. \ref{tab:q1-j-2}-b) we find out that for the first three months none of the threshold values in the set (5 and 10) satisfy R1. This shows that even though leave-based and join-based key update strategies give very similar results, there are some critical points that can affect our decision, as in this example. As a result, \emph{this strategy does not satisfy the requirement}.
\end{itemize}
{\bf Result:} At this point, we are convinced that among all three strategies only the leave-based key update strategy is convenient. Now, all we need to do is to find the exact threshold value for the given input set in Step III. As we need to make a decision between 5 and 10, we can use the efficiency graphics. In Fig. \ref{tab:q4-l-2}-b, we find out that 99,95\% of the updates will be useless when threshold is 5, and when we increase the threshold to be 10 then this percentage will only decrease to 99,90\%. Using this knowledge, we can conclude that choosing leave-based key update with threshold 10 would be the optimum solution that would balance security and also power consumption.

\fbox{\textit{\textbf{Example 2}}:} We can continue with a bit more complicated example that would also make use of the recovery results. This time, we will keep the explanations less and stress the data and findings.

{\bf Step I:} We choose the \emph{Home Automation} scenario (HA). The information that we need for model checking is given below:
\begin{itemize}
\item maximum number of devices: \textit{20 devices.}
\item average join: \emph{1 device per week.}
\item average leave: \emph{1 device per year.}
\item risk of key compromise: \emph{1\%.} 
\end{itemize}
{\bf Step II:} Then, we define our requirements.
\begin{itemize}
\item R1 - \emph{the probability that the key is compromised must be lower than \underline{10\%}, \underline{at any time}.}
\item R2 - \emph{the probability that the key recovery from a key compromise takes more than \underline{12 / 6 / 3 } months, must be lower than \underline{15 / 45 / 65 \%}.}
\item R3 - \emph{the percentage of the useless key updates should be lower than \underline{95\%}.}
\end{itemize}
{\bf Step III:} We determine the parameters for the key update strategies that satisfy the requirements. For the sake of simplicity we limit the thresholds set as \{3,6,9,12\} for the time-based key update and \{5,10,15,20\} for thee leave-based and join-based key updates. Doing so, we make it easier for the reader to verify the results from the graphs in the appendix.

To satisfy R1(we make use of the results in optimising key confidentiality): 
\begin{itemize}
  \item \textbf{Time-based: }Steady-state probabilities (Fig. \ref{tab:q2-t-1}-a) point that we can pick a threshold value below 7, and transient probabilities (Fig. \ref{tab:q1-t-2}-a) confirm this by picking 3 and 6 from the thresholds set. 
  \item \textbf{Leave-based: }Steady-state probabilities (Fig. \ref{tab:q2-l-1}-a) point that we can pick a threshold value less than or equal to 20, however transient probabilities  (Fig. \ref{tab:q1-l-2}-a) warn us not to use 20 but to use the values 5, 10, and 15 from the set.
  \item \textbf{Join-based: } Very similar to the results on the leave based strategy, we can use join-based strategy for thresholds of 5, 10, and 15 (as seen in Fig. \ref{tab:q1-j-2}-a and Fig. \ref{tab:q1-j-2}-a).
\end{itemize}
To satisfy R2 (we make use of the results in optimising recovery time):
\begin{itemize}
  \item \textbf{Time-based: }According to the result in Fig. \ref{tab:q3-t-1}-a, we can use both threshold 3 and threshold 6 to satisfy R2. 
  \item \textbf{Leave-based: }According to the result in Fig. \ref{tab:q3-l-1}-a, we can only use threshold 5 to satisfy R2.
  \item \textbf{Join-based: } According to the result in Fig. \ref{tab:q3-j-1}-a, we can only use threshold 5 to satisfy R2.
\end{itemize}
To satisfy R3 (we make use of the results in efficiency of key updates): 
\begin{itemize}
  \item \textbf{Time-based: } The threshold value of the time-based key update that satisfies R3 should be above 3, for instance threshold 6 has 91.1\% useless resets (Fig. \ref{tab:q4-t-1}-a).
  \item \textbf{Leave-based: }The threshold value of the leave-based key update that satisfies R3 should be above 5, for instance threshold 10 has 90.4\% useless resets (Fig. \ref{tab:q4-l-1}-a).
  \item \textbf{Join-based: } The threshold value of the join-based key update that satisfies R3 should be above 5, since threshold 5 has 95.1\% useless resets (Fig. \ref{tab:q4-j-1}-a).
\end{itemize}
{\bf Result:} In order to conclude on a result, we need to have the intersection of the parameter values that satisfy the constraints. For the time-based strategy, the only threshold that satisfies all of the requirements is 6. Unfortunately, the intersection of the threshold values that satisfy R1, R2, and R3 is empty set for both leave-based and join-based strategies. Therefore, the only solution for this example is the time-based key update strategy where the key is updated every three months.

\section{Related Work}
\label{sec:related}
The effect of key exchanges to the performance in an IEEE 802.15.4 (the standard that defines the physical (PHY) and medium access control (MAC) layers of ZigBee) network was recently presented in \cite{Misic:Amini}.
The authors simulated SKKE protocol (a key exchange protocol used in ZigBee, see \cite{Yuksel:Nielson:2008}) over PHY and MAC layers of a beacon enabled cluster in star topology, with strong assumptions such as all the devices had pre-installed master keys (MK, a pairwise shared key type in ZigBee) and the security suite was ENC-MIC-128 (roughly means that both \emph{ENCryption} and \emph{Message Integrity Code} security precautions are applied, and the MIC size is 128). They proposed a scheme where the coordinator maintains a counter for each node, that keeps track of number of packets exchanged under the same link key (LK, another pairwise shared key type that resembles to the notion of session key). When the threshold value of the counter is reached for any device, the coordinator initiates key exchange with all the devices. This was not a usual approach for such a network and the threshold chosen (10) was absolutely unrealistic. A less reliable approach could be single counter for all devices, but in fact both approaches suffer from the risk of DoS attack by single corrupted device. Besides, due to their star topology there was no direct connection between devices, and all the LKs were established between the device and the coordinator. This is actually fine, but against the idea between SKKE and usage of LK.

The results of \cite{Misic:Amini} indicate that even for small cluster size, frequent key exchange impose serious performance burden on data traffic; therefore period for key exchanges has to be traded for the cluster size and the throughput. Although the work presented is really interesting and new, clearly there are problems in understanding the security part (\emph{e.g.} in ZigBee-2007 SKKE does not work the way they explained, key size does not vary in ZigBee) and explanation of the formal model is missing (\emph{i.e. no detail on the Petri net simulation, except the name of the company}). Besides they only evaluated the MAC layer (throughput, blocking, etc.) and the results they achieved are no surprise.

The rest of the related work may not seem to be highly related but that is because this exact topic was not addressed before.

In \cite{Fruth:2006}, an application of probabilistic model checking to IEEE 802.15.4 is presented. Although this study is not related to security, it is a comprehensive study that developed high level generic models for contention resolution protocol used by ZigBee and evaluates performance properties. Besides, this study indicates the power of model checking approach compared to test and simulation methods by providing provably correct results that covers the full behaviour of the models.

In \cite{Yuksel:Nielson:2009}, static analysis method which is another formal method for verification is used to discover a flaw in ZigBee's key establishment protocol, and a fix is proposed. Although this study is in discrete-time domain therefore not producing quantitative but qualitative results, it is important in the sense that the usage of a formal method to ZigBee security is presented. Model checking is a similar formal method, and both of these methods have different pros and cons. In addition, key establishment is another important issue of ZigBee security and it is closely related to key update.

Finally, we would like to discuss other possible strategies for updating the keys. As we mentioned above, another approach that was proposed in \cite{Misic:Amini} is counting the number of communicated messages, therefore we call this approach \emph{message-based}. In this approach, there is a threshold value for sent messages, which triggers key update when exceeded. This approach is pretty limited in the sense that if the devices exchange messages rarely, than the keys are also updated rarely. This would cause the problem of a left or removed device to still hold a valid key, that it can share with an illegitimate device. 

We could come up with a safe proposal that we call the \textbf{hybrid} strategy, where we would make use of the previous models. Hybrid approach would try to fulfil the gap on the key update problem in the ZigBee Specification by synthesising the best of the other strategies. Depending on the network/application type we could find the best strategy for key update, using any function (\emph{e.g.} min, max, avg) on the result of the previous approaches (\emph{i.e.} time, message, join, leave). This flexibility could give us not only many choices, but also an adaptive and efficient way of key update strategy. However, this approach is not fitting with the low-rate nature of ZigBee networks since it requires the implementation of all the strategies, and definitely all the computations required by those strategies. 

\section{Conclusion}
\label{sec:conclusion}
ZigBee is an emerging wireless sensor network standard with a huge potential of being used in security-critical areas. 
Thus ZigBee networks should be able to offer the intended security guarantees.
In this paper, we have developed generic models for various application domains in ZigBee and addressed the underspecification of the key update issue.
We have presented the first application of stochastic model checking to the ZigBee security, and optimisation of key updates.
We focused on key confidentiality, key recovery time, impact of network size on the security, and the efficiency of the key updates.
We expect that due to ZigBee's wide applicability in different areas, much more work in security will be necessary.

\textbf{Future Work.} We want to improve the models so as to be more realistic:
\begin{enumerate}
\item \emph{Time-based model} - improve accuracy of timing: Current version exponentially distributes delays, we want to be more precise on the timing of the key update events in this key update strategy.
\item \emph{Leave/Join-based models} - improve instant transitions: Current versions have delays on the key update events, however those actions should be instantaneous. We want to remove unnecessary delays on reset actions.
\end{enumerate}

\clearpage

\appendix
\section{The Gap in the ZigBee Specification}
\label{app:a}
The latest ZigBee Specification (\textbf{ZigBee-2007}) \cite{ZigBee:2007}, states that ``\emph{policy decisions to \textbf{expire and periodically update keys, if desired} must be addressed correctly by any real implementation}''. In addition, it is stated that ``\emph{the \textbf{application profile}s\footnote{An \emph{Application Profile} is defined as ``\emph{a collection of device descriptions, which together form a cooperative application.}'' in the ZigBee Specification. Application profiles provide standard interfaces and device definitions to allow interoperability among ZigBee devices produced by various manufacturers, in a specific application domain.} should include these policies}''\footnote{\cite{ZigBee:2007}, Section 4.2.1.2 Security Design Choices, page 422.}.\\
\\
The latest published ZigBee application profile is the \emph{Smart Energy Application Profile} (\textbf{ZigBee-SE}) \cite{ZigBee:SE}, which is curently the most critical and important ZigBee application profile. ZigBee-SE states  that ``\emph{Periodically the trust center \textbf{shall} update the \textbf{network key} (\textbf{NK})}'' and ``\emph{Periodically the trust center \textbf{may} update the \textbf{link key} associated with a particular device.}''\footnote{\cite{ZigBee:SE}, Section 5.4 Smart Energy Profile Security, page 20. As a note on conformance levels, \emph{may} equals \emph{is permitted}, and \emph{shall} equals \emph{is required to} in the ZigBee-SE Specification.}.\\
\\
In addition to the specification documents (i.e. ZigBee-2007) and application profiles (e.g. ZigBee-SE), one should also consider the \emph{Stack Profile}s. Stack profiles are intended to support the \emph{Application Profiles}, since most of the parameters that are defined in the application profiles are set by the stack profiles. The latest published ZigBee-PRO Stack Profile \cite{ZigBeePRO:Stack} states  that ``\emph{it is \textbf{recommended} that the trust center change the network key if it is discovered that any device has been stolen or otherwise compromised, ...}'' and ``\emph{there is \textbf{no expectation} that the network key be changed when \textbf{adding a new device}.}''.\\
\\
As seen clearly from our explanations with precise references above, all the specification documents of ZigBee leave the important key update issue to the implementations.\\

\clearpage
\section{Model Checking Results for All Application Profiles}
\label{app:b}
\input{experiments}

\end{document}

%% file: title.tex

\begin{titlepage}

\begin{center}

\vspace{25mm}
{\huge \emph{Optimizing ZigBee Security\\ using Stochastic Model Checking}}

\vspace{35mm}

	\emph{Ender Y\"uksel}  \hspace{10mm}  \emph{Hanne Riis Nielson} \hspace{10mm}  \emph{Flemming Nielson} \\
\vspace{3mm}
	Technical University of Denmark \\
  Informatics and Mathematical Modelling \\ 
  \tt{\{ey,riis,nielson\}@imm.dtu.dk}

\vspace{10mm}

\textnormal{	\emph{Matthias Fruth}  \hspace{10mm}  \emph{Marta Kwiatkowska}  }\\
\vspace{3mm}
\textnormal{	Oxford University }\\
\textnormal{  Computing Laboratory }\\
  \tt{\{matthias.fruth,marta.kwiatkowska\}@comlab.ox.ac.uk}

\vspace{4cm}

\large

\textnormal{Technical University of Denmark}

\vspace{5mm}

\textnormal{IMM-TECHNICAL REPORT-2010-08}

\vspace{1cm}

\textnormal{ \today }

\end{center}

\end{titlepage}

%% file: colophon.tex
\ \\

\vspace{15cm}

\ \\
Technical University of Denmark\\
Informatics and Mathematical Modelling\\
Building 321, DK-2800 Lyngby, Denmark\\
Phone +45 45253351, Fax +45 45882673\\
reception@imm.dtu.dk\\
www.imm.dtu.dk\\
\ \\
IMM-TECHNICAL REPORT: ISSN 1601-2321

\thispagestyle{empty}

\newpage

%% file: experiments.tex
In this section, we have a systematic and complete set of probabilistic model checking experiments, where we explore 3 key update strategies, on 6 application scenarios by asking 4 different question each time; thus we present 72 (3*6*4) graphics, also known as experiments in probabilistic model checking jargon. Among these 72 experiments, 

\begin{itemize}
\item 12 experiments were conducted for 120 months and 4 different threshold values, yielding 5760 model checkings
\item 12 experiments were conducted for 60 months and 4 different threshold values, yielding 2880 model checkings
\item 12 experiments were conducted for 24 months and 4 different threshold values, yielding 1152 model checkings
\item 7 experiments were conducted for 8 different threshold values, yielding 56 model checkings
\item 7 experiments were conducted for 4 different threshold values, yielding 28 model checkings
\item 4 experiments were conducted for 40 different threshold values, yielding 160 model checkings
\item 4 experiments were conducted for 20 different threshold values, yielding 80 model checkings
\item 2 experiments were conducted for 80 different threshold values, yielding 160 model checkings
\item 2 experiments were conducted for 48 different threshold values, yielding 96 model checkings
\item 2 experiments were conducted for 40 different threshold values, yielding 80 model checkings
\item 2 experiments were conducted for 24 different threshold values, yielding 48 model checkings
\item 2 experiments were conducted for 10 different threshold values, yielding 20 model checkings
\item 2 experiments were conducted for 5 different threshold values, yielding 10 model checkings
\item 1 experiments were conducted for 96 different threshold values, yielding 96 model checkings
\item 1 experiments were conducted for 12 different threshold values, yielding 12 model checkings
\end{itemize}
In total, 10638 model checkings were conducted.

In some application scenarios, the number of devices in the network were too large, therefore requiring adjustments in the \emph{convergence}.  Specifically, to decide the termination of the iterative methods that are used for computation of probabilities, we have parameters such as termination epsilon, termination criteria, and maximum number of iterations. For regular experiments, we used the Power Method with a maximum of 10000 iterations. However, for the cases that this was not enough for convergence, we chose Pseudo Gauss-Seidel method.

\subsection{The Application Scenarios}
\label{app:sce}
In this section, we present the details of the remaining three application profiles, that were left to the appendix. As we have done for the \emph{Home Automation}, \emph{Smart Energy}, and \emph{Commercial Building Automation} scenarios, this time we will present the technical details in terms of maximum network size, rate of join, rate of leave, and key compromise probability for the scenarios \emph{Personal, Home and Hospital Care}, \emph{Telecom Applications}, and \emph{Wireless Sensor Applications}.

\paragraph{The Personal, Home and Hospital Care Scenario:} 
In this scenario, the network is \emph{dynamic} such that it is likely that devices join and leave the network regularly. The network size is in general less than 1000 devices. The environment is very secure, still the key updates are needed to be rather frequent. Based on these assumptions we specify the remaining constants as follows:
\begin{quote}\small
\begin{verbatim}
const int    Max      = 500;    // maximal size of network 
const double R_leave  = 1/30; // average leave: 1 per month
const double P_comp   = 1/10000; // risk of key compromise 
\end{verbatim}
\end{quote}

\paragraph{The Telecom Applications Scenario:} 
In this scenario, the network is \emph{dynamic} such that the nodes are joining and leaving frequently. The network size is in general less than 50 devices. The environment is highly secure, therefore key updates are needed to be rather infrequent. Based on these assumptions we specify the remaining constants as follows:
\begin{quote}\small
\begin{verbatim}
const int    Max      = 20;    // maximal size of network 
const double R_leave  = 1/30; // average leave: 1 per month
const double P_comp   = 1/100000; // risk of key compromise 
\end{verbatim}
\end{quote}

\paragraph{The Wireless Sensor Applications Scenario:} 
In this scenario, the network is \emph{fairly dynamic} though it depends on the particular application. The network size is in general less than 1000 devices. The environment is secure, still the key updates are needed to be rather frequent. Based on these assumptions we specify the remaining constants as follows:
\begin{quote}\small
\begin{verbatim}
const int    Max      = 500;    // maximal size of network 
const double R_leave  = 1/(6*30); // average leave: 1 per 6 months
const double P_comp   = 1/1000; // risk of key compromise 
\end{verbatim}
\end{quote}
\subsection{The Key Update Thresholds}
\label{app:kuthresh}
In this section, we present the values of the key update thresholds that we used in the experiment. In Table \ref{tab:thresholds}, we present the threshold values in \texttt{start - end - step} format. For example, in the experiments regarding the \emph{Home Automation} scenario which employs the \emph{Time-Based Key Update} strategy include four different threshold values for Question 1. These values are: 3 ({\tt start}), 6 ({\tt start} + {\tt step}), 9 (previous + {\tt step}), and 12 ({\tt end}).

\begin{table}[!h]
\centering
\caption{Key Update Thresholds}
\label{tab:thresholds}
\begin{tabular}{c|c|cccc|c} \hline
\textbf{Scenario} &\textbf{Strategy} & \textbf{Question 1} & \textbf{Question 2} & \textbf{Question 3} & \textbf{Question 4} & \textbf{Duration}\\ \hline
\multirow{3}{*}{\textbf{HA}} & TB & 3 - 12 - 3 & 1 - 12 - 1 & 3 - 12 - 3 & 1 - 12 - 1& \multirow{3}{*}{60 months} \\
                                                & LB & 5 - 20 - 5 & 1 - 20 - 1 & 5 - 20 - 5 & 1 - 20 - 1\\
                                                & JB & 5 - 20 - 5 & 1 - 20 - 1 & 5 - 20 - 5 & 1 - 20 - 1\\ \hline
\multirow{3}{*}{\textbf{SE}} & TB & 12 - 48 - 12 & 1 - 48 - 1 & 12 - 48 - 12 & 1 - 48 - 1 & \multirow{3}{*}{120 months} \\
                                                & LB & 2 - 5 - 1 & 1 - 5 - 1 & 2 - 5 - 1 & 1 - 5 - 1\\
                                                & JB & 2 - 5 - 1 & 1 - 5 - 1 & 2 - 5 - 1 & 1 - 5 - 1\\ \hline
\multirow{3}{*}{\textbf{CBA}} & TB & 6 - 24 - 6 & 1 - 24 - 1 & 6 - 24 - 6 & 1 - 24 - 1 & \multirow{3}{*}{120 months} \\
                                                & LB & 10 - 40 - 10 & 1 - 40 - 1 & 10 - 40 - 10 & 1 - 40 - 1\\
                                                & JB & 10 - 40 - 10 & 1 - 40 - 1 & 10 - 40 - 10 & 1 - 40 - 1\\ \hline
\multirow{3}{*}{\textbf{PHHC}} & TB & 1 - 4 - 1 & 1 - 4 - 1 & 1 - 4 - 1 & 1 - 4 - 1 & \multirow{3}{*}{24 months} \\
                                                & LB & 5 - 20 - 5 & 5 - 20 - 5 & 5 - 20 - 5 & 5 - 20 - 5\\
                                                & JB & 5 - 20 - 5 & 5 - 20 - 5 & 5 - 20 - 5 & 5 - 20 - 5\\ \hline
\multirow{3}{*}{\textbf{TA}} & TB & 1 - 4 - 1 & 1 - 4 - 1 & 1 - 4 - 1 & 1 - 4 - 1 & \multirow{3}{*}{60 months} \\
                                                & LB & 5 - 20 - 5 & 1 - 20 - 1 & 5 - 20 - 5 & 1 - 20 - 1\\
                                                & JB & 5 - 20 - 5 & 1 - 20 - 1 & 5 - 20 - 5 & 1 - 20 - 1\\ \hline
\multirow{3}{*}{\textbf{WSA}} & TB & 1 - 4 - 1 & 1 - 4 - 1 & 1 - 4 - 1 & 1 - 4 - 1 & \multirow{3}{*}{24 months} \\
                                                & LB & 5 - 20 - 5 & 5 - 20 - 5 & 5 - 20 - 5 & 5 - 20 - 5\\
                                                & JB & 5 - 20 - 5 & 5 - 20 - 5 & 5 - 20 - 5 & 5 - 20 - 5\\ \hline
\end{tabular}
\end{table}

\subsection{The Questions} 
\label{sec:questions}

In this section, we summarize the questions that we are interested in. We specify these questions in CSL and present the answers for those questions with graphics in the following sections.

\subsubsection{Question 1: Key confidentiality}
\begin{quote}
What is the probability that the key is compromised at month {\tt T}?
\begin{quote}
\begin{verbatim}
P=? [ F[30*T,30*T] Comp ]
\end{verbatim}
\end{quote}
\end{quote}
The formula expresses the probability of being in a state where the key is compromised at time instant {\tt 30*T}, that is, when exactly {\tt T} months have passed. 

\subsubsection{Question 2: Long run behaviour}
\begin{quote}
What is the steady state probability of the key being compromised?
\begin{quote}
\begin{verbatim}
S=? [Comp]
\end{verbatim}
\end{quote}
\end{quote}
The formula expresses the long run behaviour of key confidentiality for the given model.

\subsubsection{Question 3: Key recovery}
\begin{quote}
What is the risk that it takes \emph{more than} {\tt T} months to recover from a compromise of the key?
\begin{quote}
\begin{verbatim}
P=? [Comp U>=(30*T) !Comp {Comp}{max}]
\end{verbatim}
\end{quote}
\end{quote}
The formula {\tt Comp U>=(30*T) !Comp} determines those paths where the key is compromised for \emph{at least} {\tt T} months before it is reset. In PRISM the filter {\tt \{Comp\}} means that we restrict our attention to those paths that start in a state where the key is compromised and {\tt \{max\}} means that we take the maximal probability of encountering one of the desired paths when starting in one of the states where {\tt Comp} holds. 

\subsubsection{Question 4: Efficiency}
\begin{quote}
In the long run, what will be the percentages of the useful resets and useless resets. A reset is useful when the key was compromised, and useless (in terms of power consumption) when the key was not compromised.
\begin{quote}
\begin{verbatim}
(100 * R{"Useful_Resets"}=? [ S ]) / R{"All_Resets"}=? [ S ] 
(100 * R{"Useless_Resets"}=? [ S ]) / R{"All_Resets"}=? [ S ] 
\end{verbatim}
\end{quote}
where the rewards {\tt All\_Reset}, {\tt Useless\_Reset} and {\tt Useful\_Reset} are given by:
\begin{quote}
\begin{verbatim}
rewards "All_Resets"
  [reset] true:  1;
endrewards

rewards "Useful_Resets"
  [reset] Comp:  1;
endrewards

rewards "Useless_Resets"
  [reset] !Comp:  1;
endrewards
\end{verbatim}
\end{quote}
\end{quote}

\subsection{The Answers} 
\label{app:answers}

In this section, we present all the results of the stochastic model checking experiments that try to answer the questions we have explained before. In parallel to the questions, this section is organised in four subsections where each section corresponds to a question. Every subsection has three pages of graphical results, each of which correspond to a unique key update strategy.  Every graphical result page is composed of two figures, first of which has two different graphics that is HA and SE application profiles, and the second of which has four different graphics covering the remaining four application profiles CBA, PHHC, TA, and WSA. Each graphic uses different colours to make it easier to distinguish between different parameter values. The number of different parameter values in a graphic is limited to 4 in order to increase the readability. 

All of the experiments can easily be replicated, to do that the models in Section \ref{sec:models}, the logic formulae in Appendix \ref{sec:questions}, the scenario inputs in Section \ref{sec:appsce} and Appendix \ref{app:sce}, and finally the key update threshold values in Appendix \ref{app:kuthresh}  should be used.
\subsubsection{Key Confidentiality}
\begin{center} 
\begin{figure}[!h]
  \caption{Results of question 1 for HA and SE in the time based environment}
    \begin{tabular}{cc}
      \includegraphics[width=8.5cm]{q1-ha-tb} & \includegraphics[width=8.5cm]{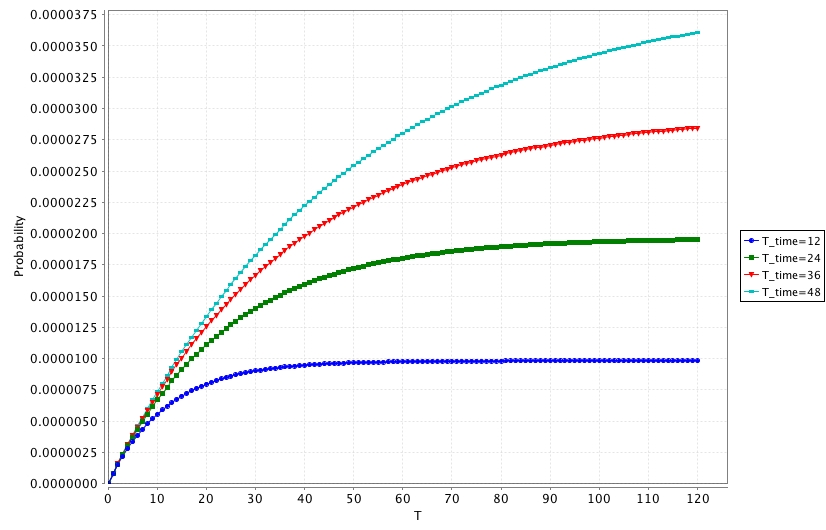}       \\
      \small{a) Home Automation} & \small{b) Smart Energy} \\
    \end{tabular}
  \label{tab:q1-t-1}
\end{figure}
\end{center}

\begin{center} 
\begin{figure}[!h]
  \caption{Results of question 1 for CBA, PHHC, TA, and WSA in the time based environment}
    \begin{tabular}{cc}
      \includegraphics[width=8.5cm]{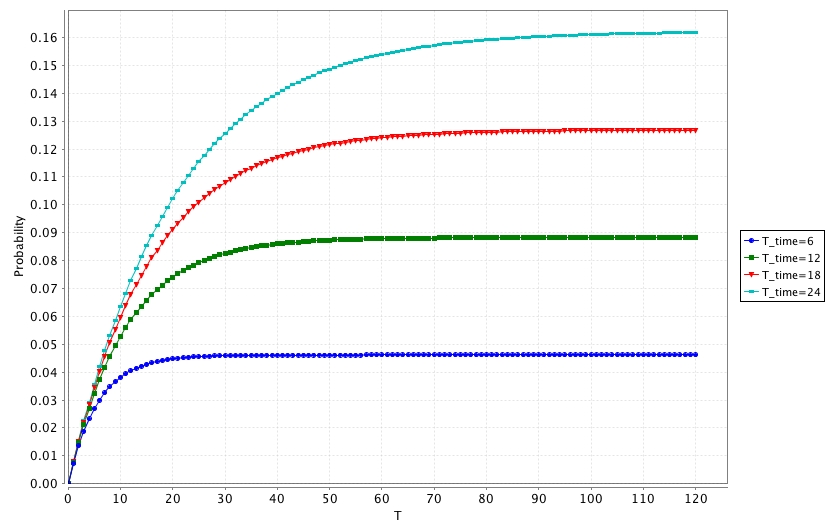} & \includegraphics[width=8.5cm]{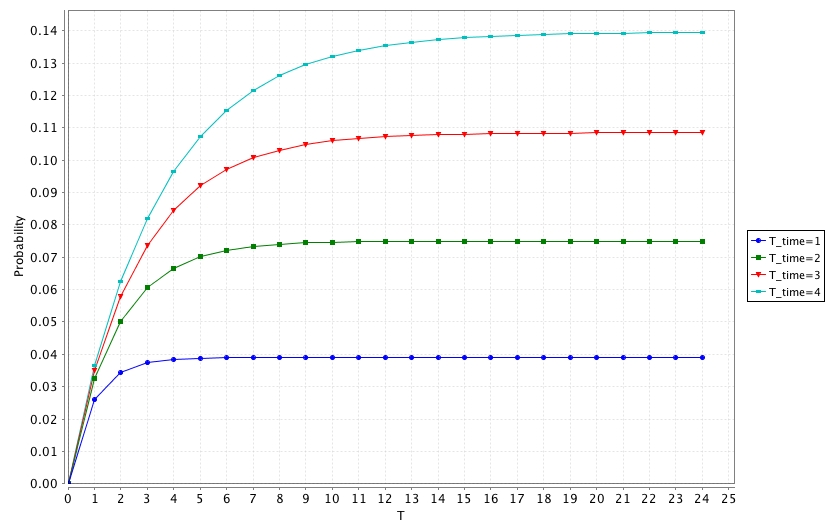}       \\
      \small{a) Commercial Building Automation} & \small{b) Personal, Home and Hospital Care} \\
      \includegraphics[width=8.5cm]{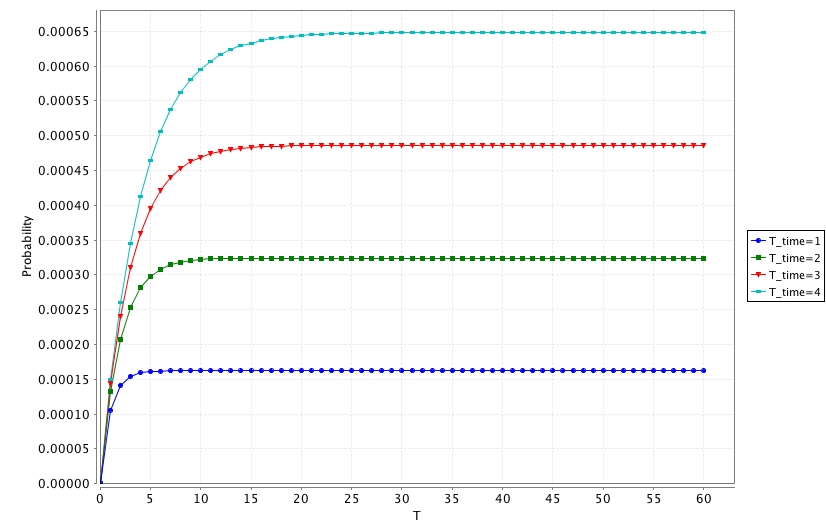} & \includegraphics[width=8.5cm]{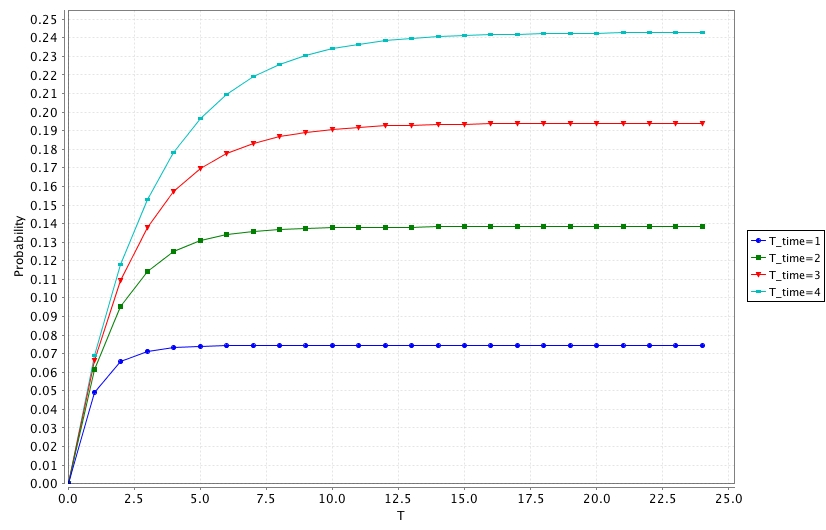}       \\
      \small{c) Telecom Applications} & \small{d) Wireless Sensor Applications} \\
    \end{tabular}
  \label{tab:q1-t-2}
\end{figure}
\end{center}

\begin{center} 
\begin{figure}[!h]
  \caption{Results of question 1 for HA and SE in the leave based environment}
    \begin{tabular}{cc}
      \includegraphics[width=8.5cm]{q1-ha-lb} & \includegraphics[width=8.5cm]{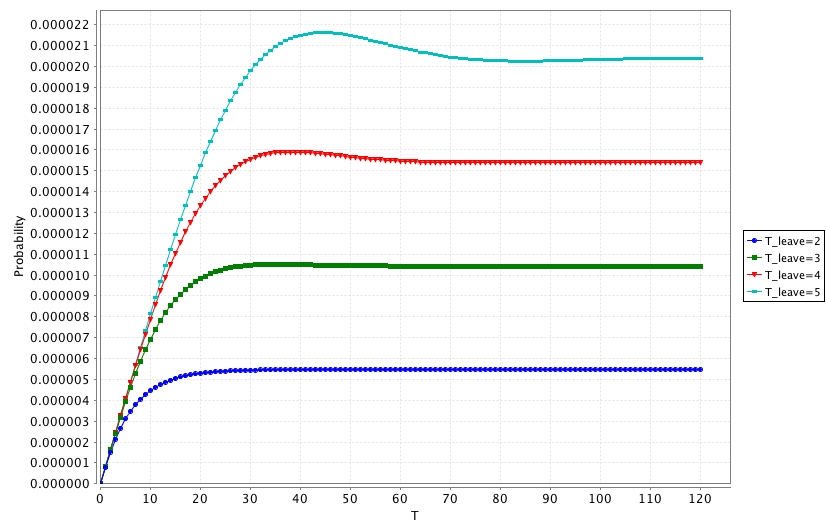}       \\
      \small{a) Home Automation} & \small{b) Smart Energy} \\
    \end{tabular}
  \label{tab:q1-l-1}
\end{figure}
\end{center}

\begin{center} 
\begin{figure}[!h]
  \caption{Results of question 1 for CBA, PHHC, TA, and WSA in the leave based environment}
    \begin{tabular}{cc}
      \includegraphics[width=8.5cm]{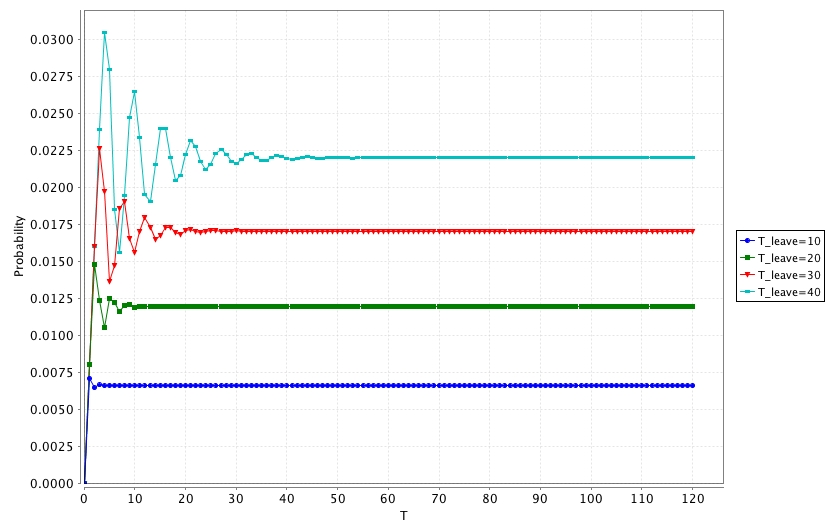} & \includegraphics[width=8.5cm]{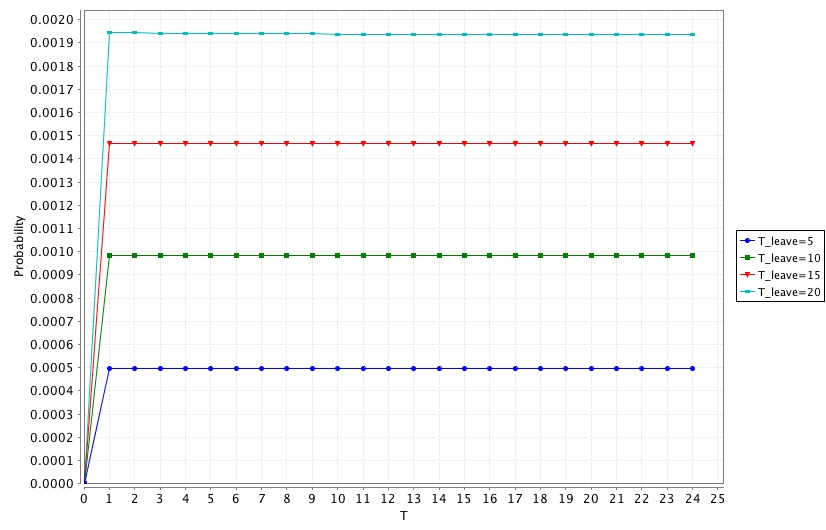}       \\
      \small{a) Commercial Building Automation} & \small{b) Personal, Home and Hospital Care} \\
      \includegraphics[width=8.5cm]{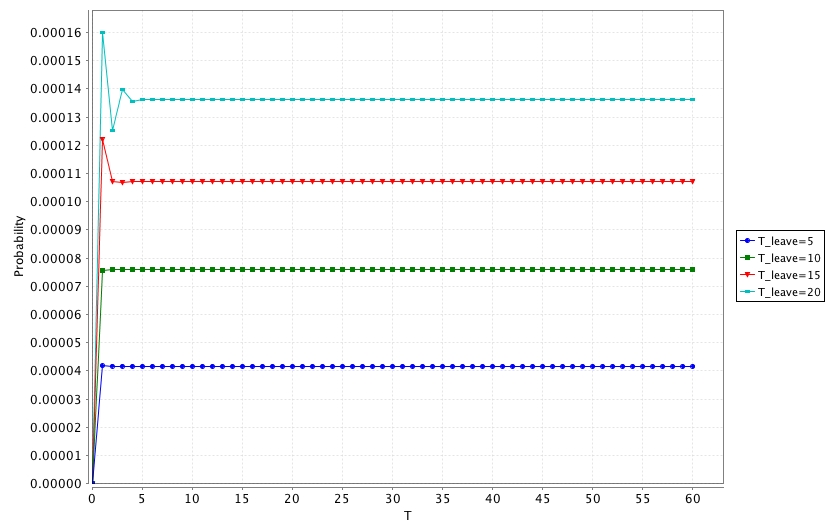} & \includegraphics[width=8.5cm]{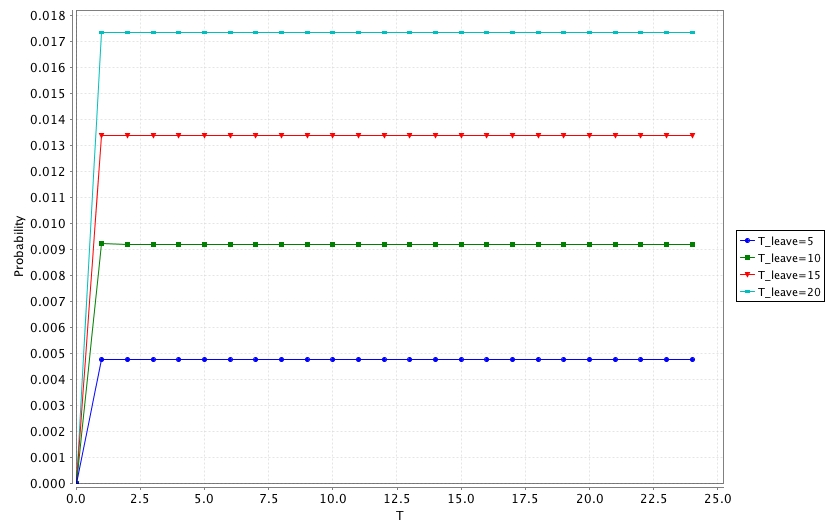}       \\
      \small{c) Telecom Applications} & \small{d) Wireless Sensor Applications} \\
    \end{tabular}
  \label{tab:q1-l-2}
\end{figure}
\end{center}

\begin{center} 
\begin{figure}[!h]
  \caption{Results of question 1 for HA and SE in the join based environment}
    \begin{tabular}{cc}
      \includegraphics[width=8.5cm]{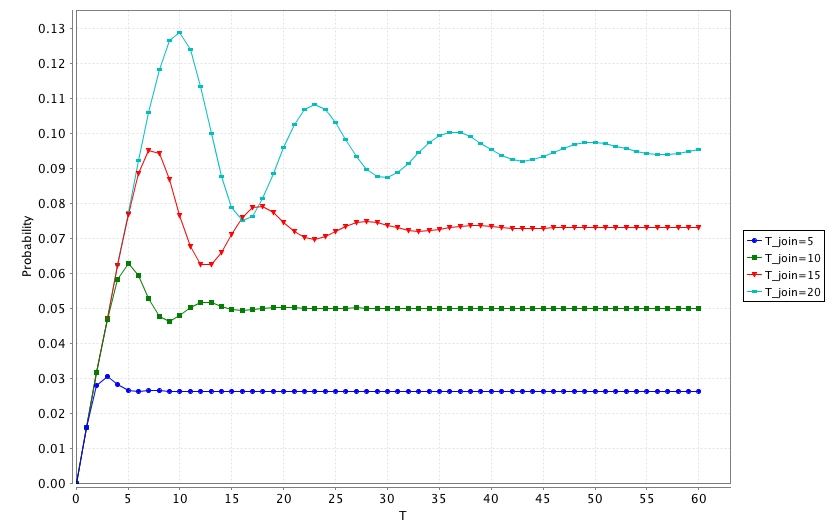} & \includegraphics[width=8.5cm]{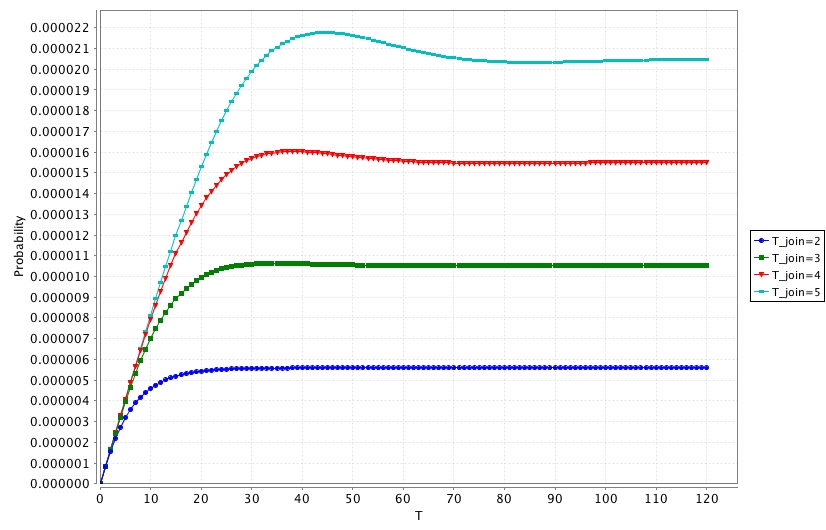}       \\
      \small{a) Home Automation} & \small{b) Smart Energy} \\
    \end{tabular}
  \label{tab:q1-j-1}
\end{figure}
\end{center}

\begin{center} 
\begin{figure}[!h]
  \caption{Results of question 1 for CBA, PHHC, TA, and WSA in the join based environment}
    \begin{tabular}{cc}
      \includegraphics[width=8.5cm]{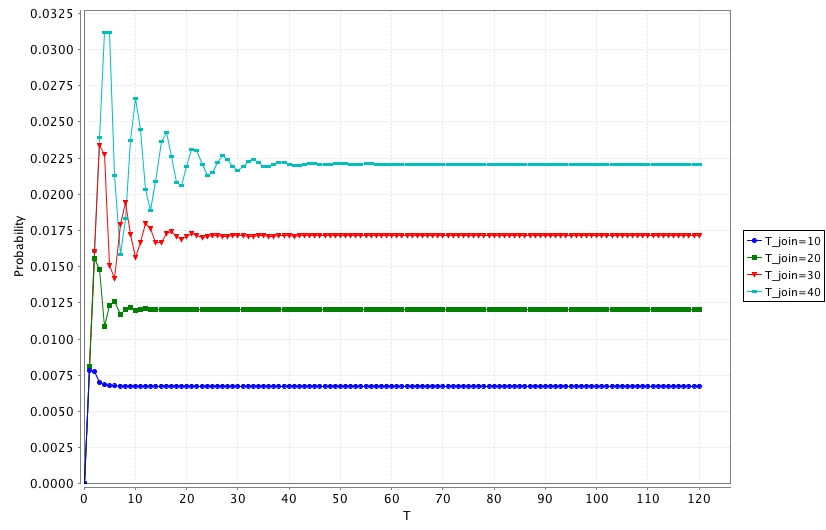} & \includegraphics[width=8.5cm]{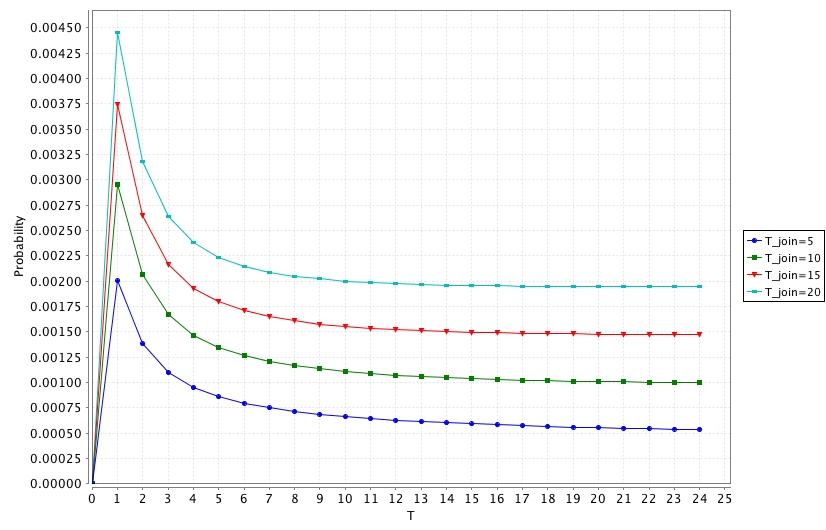} \\
      \small{a) Commercial Building Automation} & \small{b) Personal, Home and Hospital Care} \\
      \includegraphics[width=8.5cm]{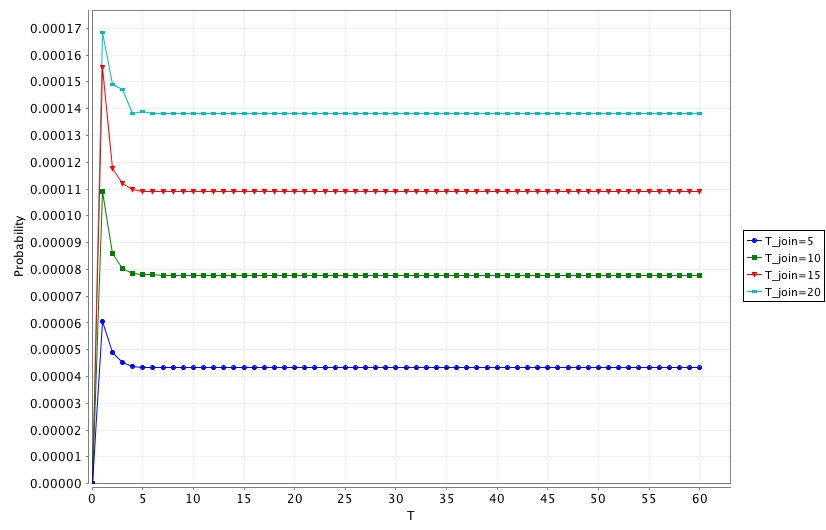} & \includegraphics[width=8.5cm]{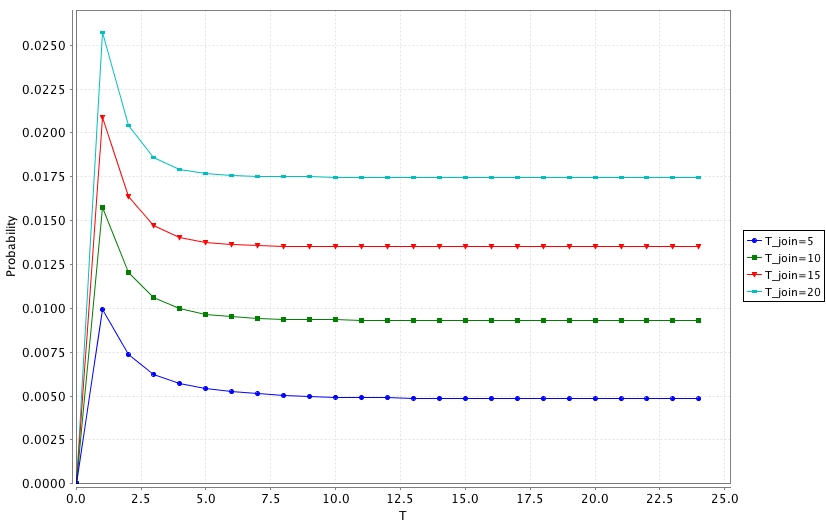} \\
      \small{c) Telecom Applications} & \small{d) Wireless Sensor Applications} \\
    \end{tabular}
  \label{tab:q1-j-2}
\end{figure}
\end{center}

Fig. \ref{tab:q1-t-1} and Fig. \ref{tab:q1-t-2} show the results for the \emph{time-based environment}, 
Fig. \ref{tab:q1-l-1} and Fig. \ref{tab:q1-l-2} show the results for the \emph{leave-based environment}, 
and Fig. \ref{tab:q1-j-1} and Fig. \ref{tab:q1-j-2} show the results for the \emph{join-based environment}.

The odd numbered figures contains the results for the scenarios HA, and SE; 
whereas the even numbered figures contains the results for the scenarios CBA, PHHC, TA, and WSA.

The x-axis always shows the time, specifically month {\tt T} of interest. 
On the y-axis we have the probability (or risk) that the key is compromised at a time instant.  
The curves represent different intervals {\tt T\_time} for resetting the key, depending on the selected scenario. 
For example, a curve labeled as {\tt T\_time=6} means that the key is updated every 6 months, periodically. 

In \textbf{time-based key update}, there is always an initial phase where the key compromise probability increases as the time passes. After that phase, the probability keeps constant for the remaining time period. This behaviour of the systems, is independent of the selection of the scenario nor the key update threshold value. We can briefly say that key compromise probability stabilizes eventually, and how long it takes to stabilize depends on 1) the key update scenario, and then 2) the threshold value.

In \textbf{leave-based key update}, there is also an initial phase where the key compromise probability increases as the time passes. However, after that phase, there is another phase where the key update probability oscillates until the probability gets stabilized. The oscillation, more specifically the \emph{peak-to-peak amplitude} is monotonically decreasing in this phase between the initial and the stabilised phases. Still, we should note that in specific cases (e.g. PHHC and WSA scenarios) where the number of devices is huge (e.g. 500 in Fig. \ref{tab:q1-l-2}-b and Fig. \ref{tab:q1-l-2}-d), the oscillations are not present or negligible. In addition, in other specific cases (e.g. SE scenario) where the number of devices is too few (e.g. 5 in Fig. \ref{tab:q1-l-1}-b), the oscillations have very low amplitude.

In \textbf{join-based key update}, is pretty much similar to the leave-based key update. So let us just talk about the differences and the unique parts of the behaviour. Only in the specific cases that the number of devices is too many (e.g. PHHC and WSA scenarios), instead of the oscillation there exists a major drop in the key compromise probability. In other words, the probability tends to increase monotonically as time passes in the beginning, and after reaching a maximum point it starts to decrease monotonically until it stabilizes (see Fig. \ref{tab:q1-j-2}-b and Fig. \ref{tab:q1-j-2}-d). 

\clearpage
\pagebreak

\clearpage
\pagebreak

\clearpage
\pagebreak

\clearpage
\pagebreak
\subsubsection{Long Run Behaviour}
\begin{center} 
\begin{figure}[!h]
  \caption{Results of question 2 for HA and SE in the time based environment}
    \begin{tabular}{cc}
      \includegraphics[width=8.5cm]{q2-ha-tb} & \includegraphics[width=8.5cm]{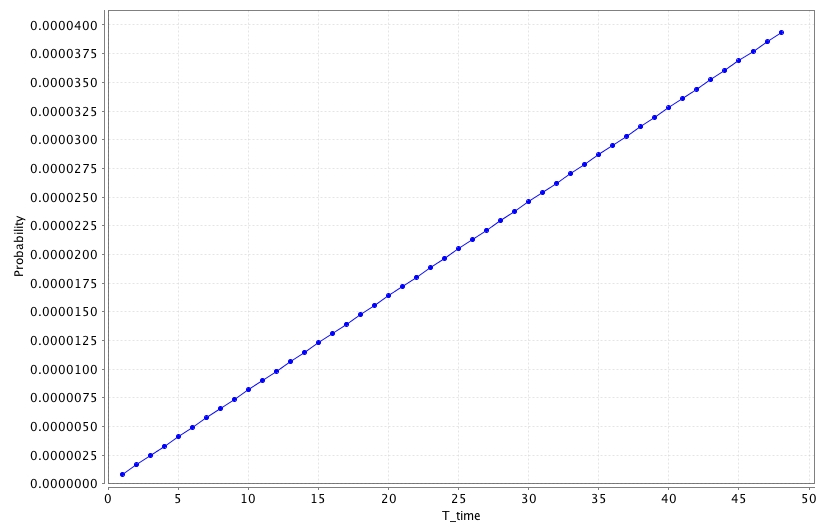}       \\
      \small{a) Home Automation} & \small{b) Smart Energy} \\
    \end{tabular}
  \label{tab:q2-t-1}
\end{figure}
\end{center}

\begin{center} 
\begin{figure}[!h]
  \caption{Results of question 2 for CBA, PHHC, TA, and WSA  in the time based environment}
    \begin{tabular}{cc}
      \includegraphics[width=8.5cm]{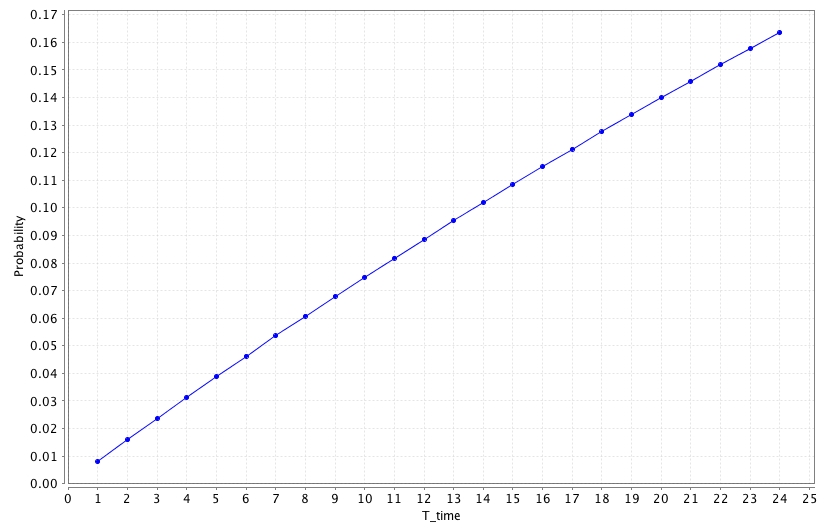} & \includegraphics[width=8.5cm]{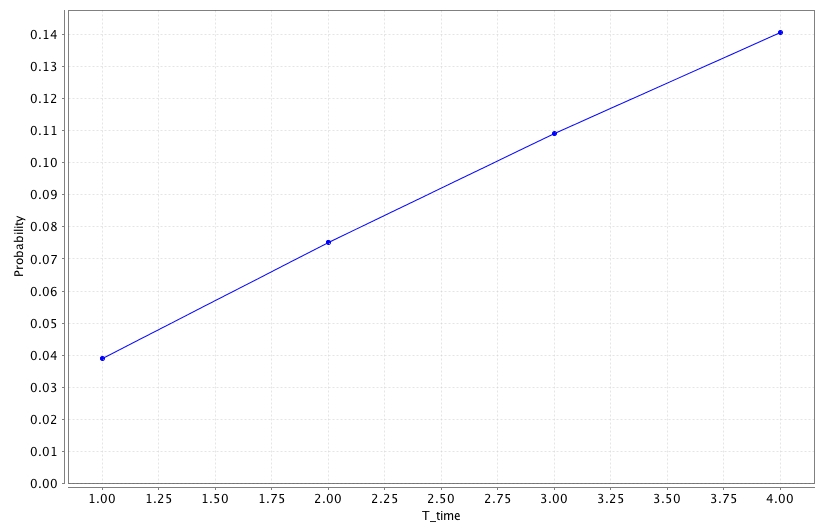}       \\
      \small{a) Commercial Building Automation} & \small{b) Personal, Home and Hospital Care} \\
      \includegraphics[width=8.5cm]{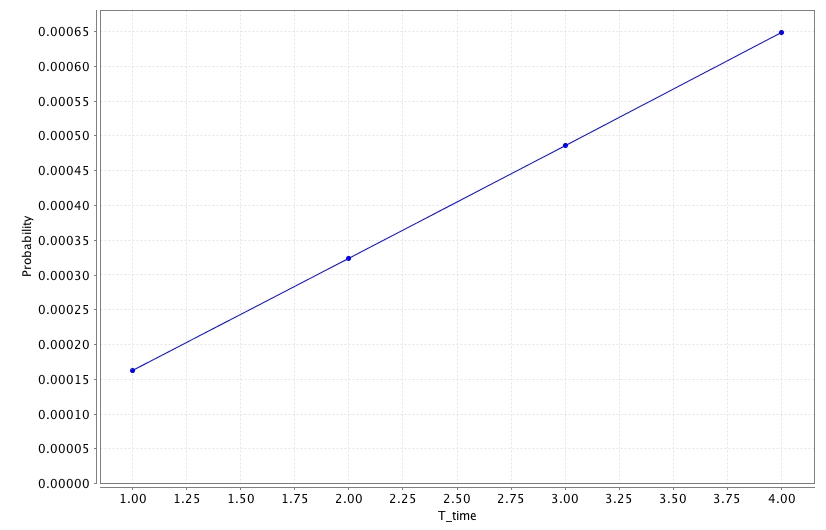} & \includegraphics[width=8.5cm]{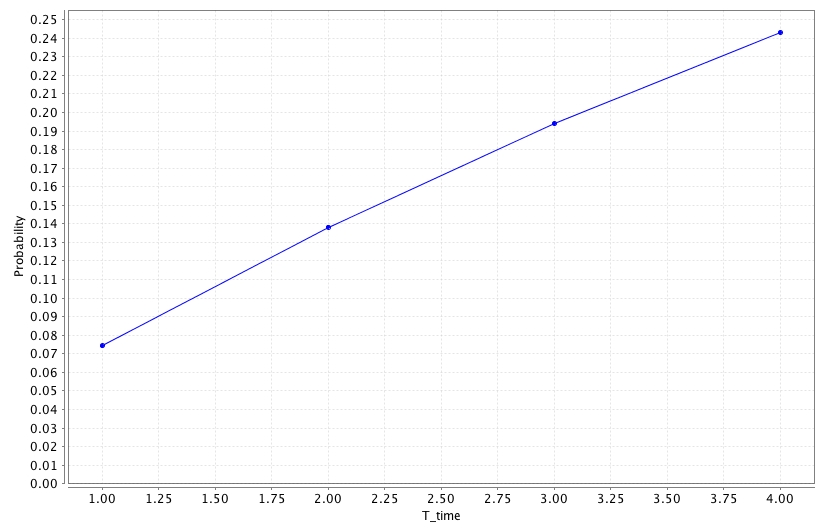}       \\
      \small{c) Telecom Applications} & \small{d) Wireless Sensor Applications} \\
    \end{tabular}
  \label{tab:q2-t-2}
\end{figure}
\end{center}

\begin{center} 
\begin{figure}[!h]
  \caption{Results of question 2 for HA and SE in the leave based environment}
    \begin{tabular}{cc}
      \includegraphics[width=8.5cm]{q2-ha-lb} & \includegraphics[width=8.5cm]{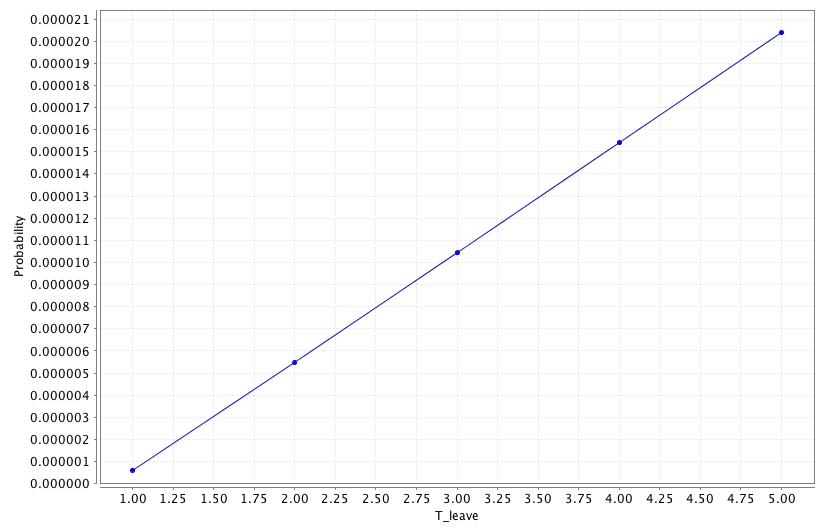}       \\
      \small{a) Home Automation} & \small{b) Smart Energy} \\
    \end{tabular}
  \label{tab:q2-l-1}
\end{figure}
\end{center}

\begin{center} 
\begin{figure}[!h]
  \caption{Results of question 2 for CBA, PHHC, TA, and WSA in the leave based environment}
    \begin{tabular}{cc}
      \includegraphics[width=8.5cm]{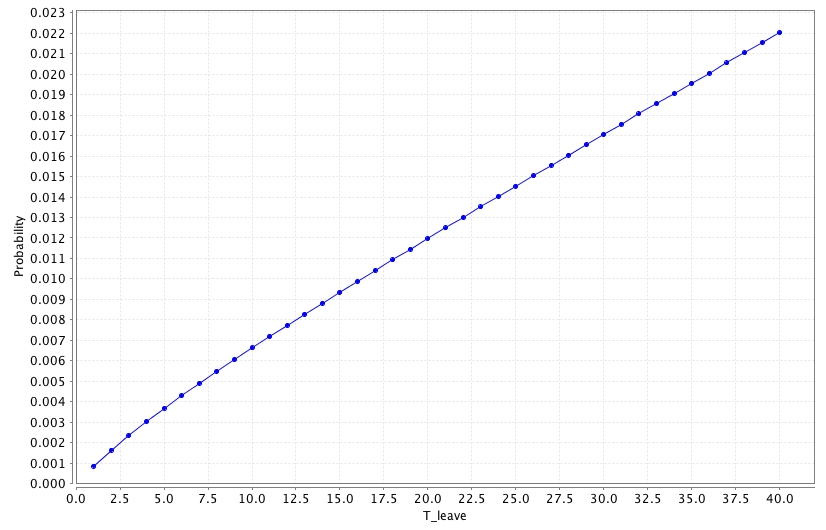} & \includegraphics[width=8.5cm]{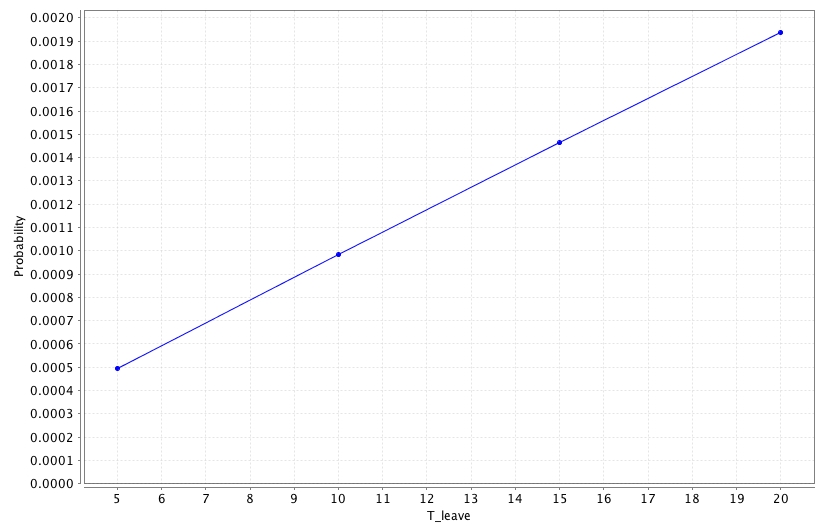}       \\
      \small{a) Commercial Building Automation} & \small{b) Personal, Home and Hospital Care} \\
      \includegraphics[width=8.5cm]{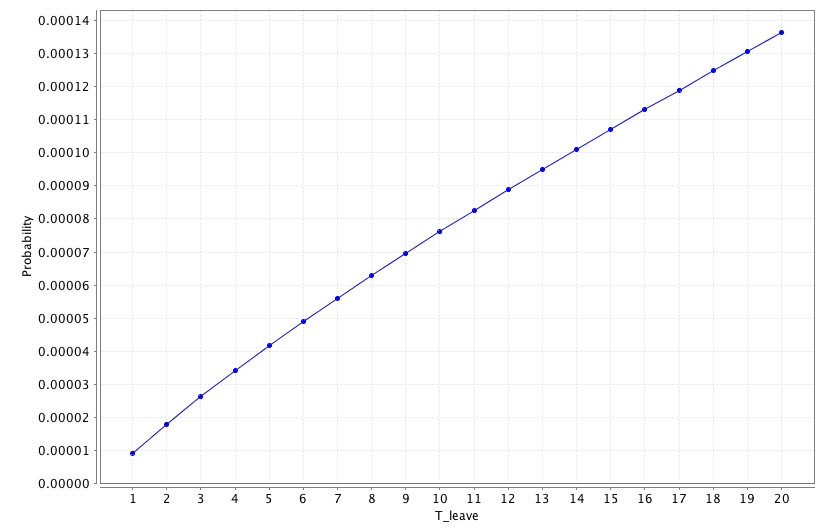} & \includegraphics[width=8.5cm]{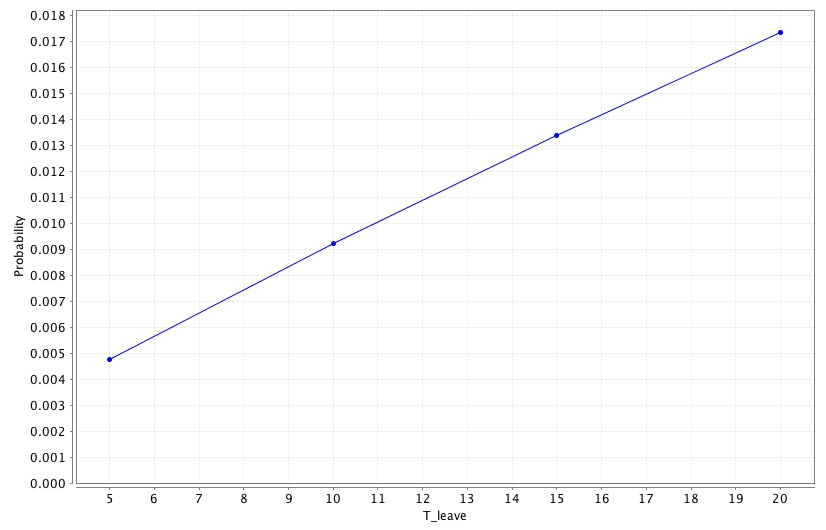}       \\
      \small{c) Telecom Applications} & \small{d) Wireless Sensor Applications} \\
    \end{tabular}
  \label{tab:q2-l-2}
\end{figure}
\end{center}

\begin{center} 
\begin{figure}[!h]
  \caption{Results of question 2 for HA and SE in the join based environment}
    \begin{tabular}{cc}
      \includegraphics[width=8.5cm]{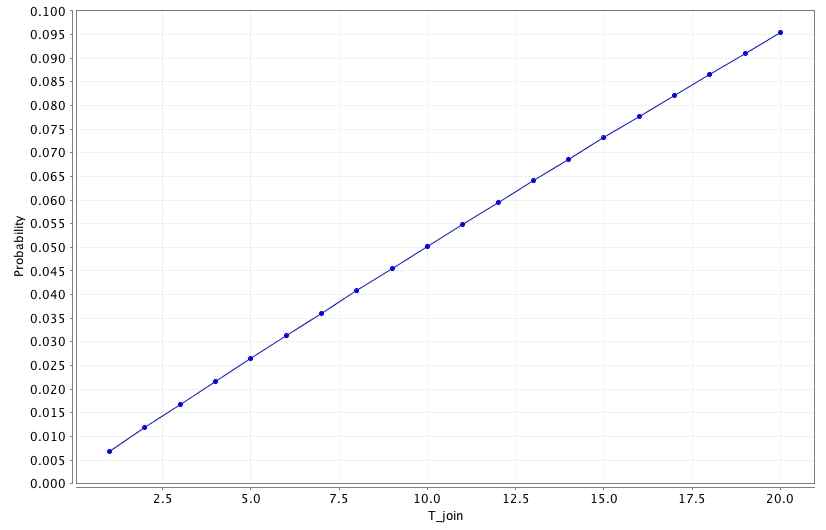} & \includegraphics[width=8.5cm]{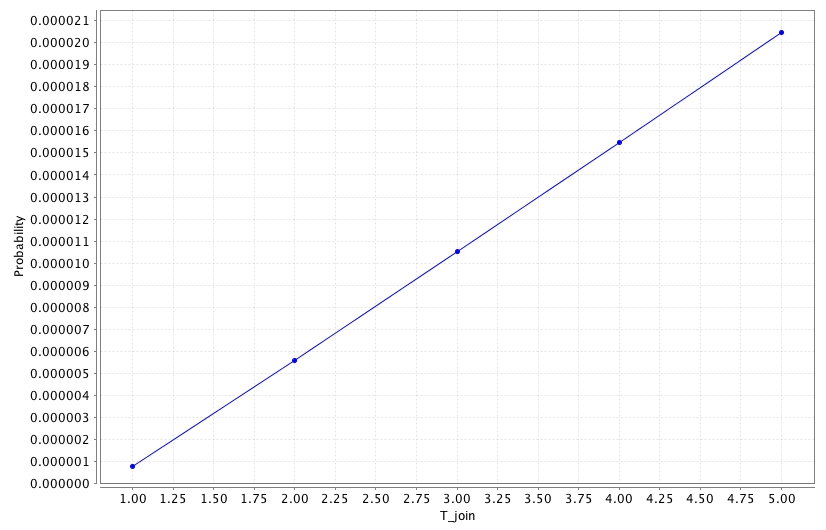}       \\
      \small{a) Home Automation} & \small{b) Smart Energy} \\
    \end{tabular}
  \label{tab:q2-j-1}
\end{figure}
\end{center}

\begin{center} 
\begin{figure}[!h]
  \caption{Results of question 2 for CBA, PHHC, TA, and WSA in the join based environment}
    \begin{tabular}{cc}
      \includegraphics[width=8.5cm]{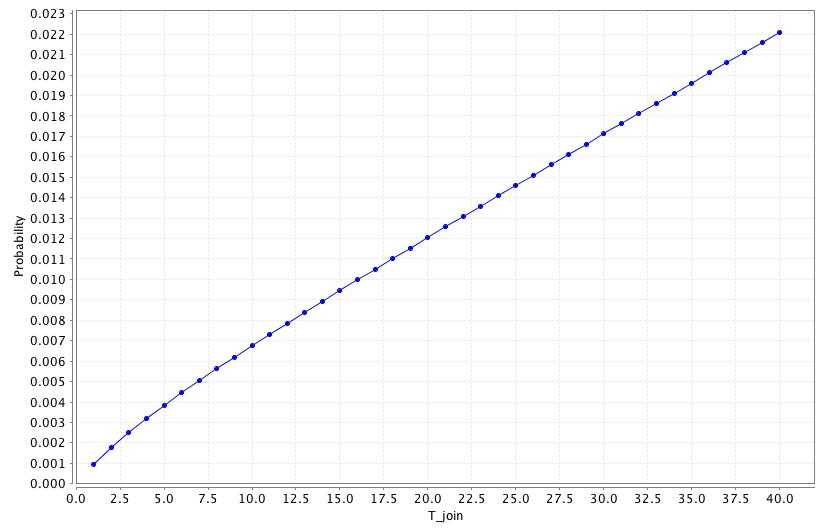} & \includegraphics[width=8.5cm]{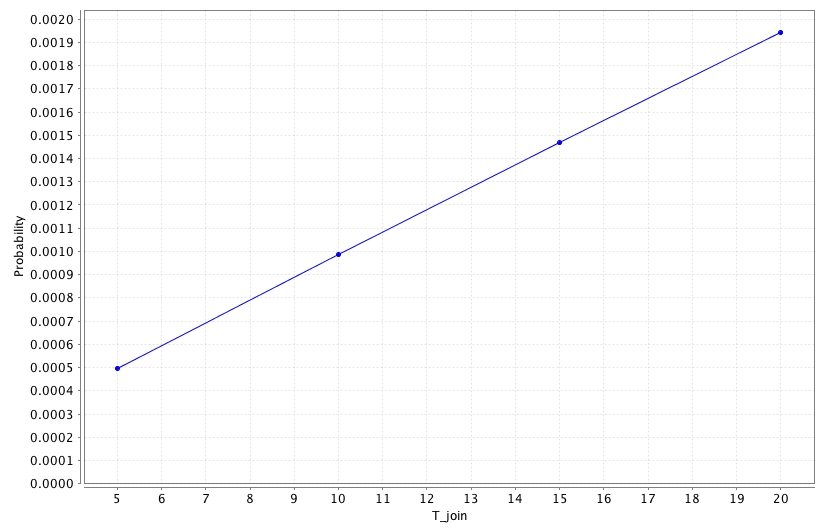} \\
      \small{a) Commercial Building Automation} & \small{b) Personal, Home and Hospital Care} \\
     \includegraphics[width=8.5cm]{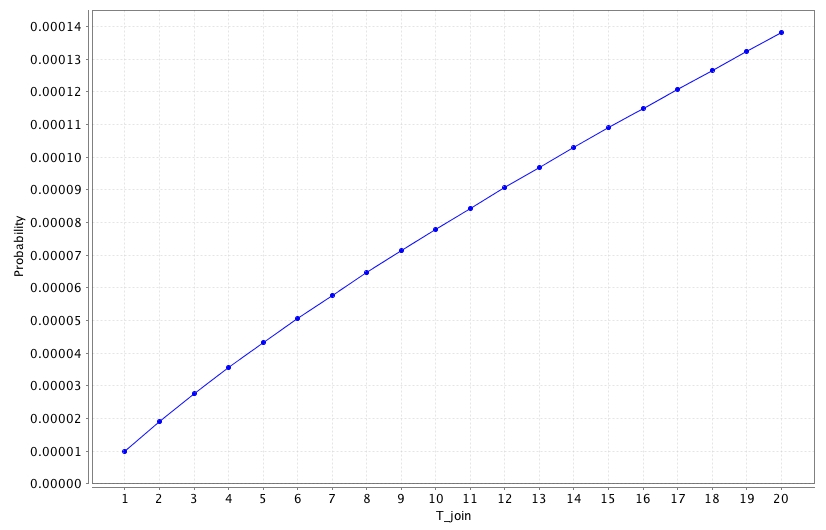}  & \includegraphics[width=8.5cm]{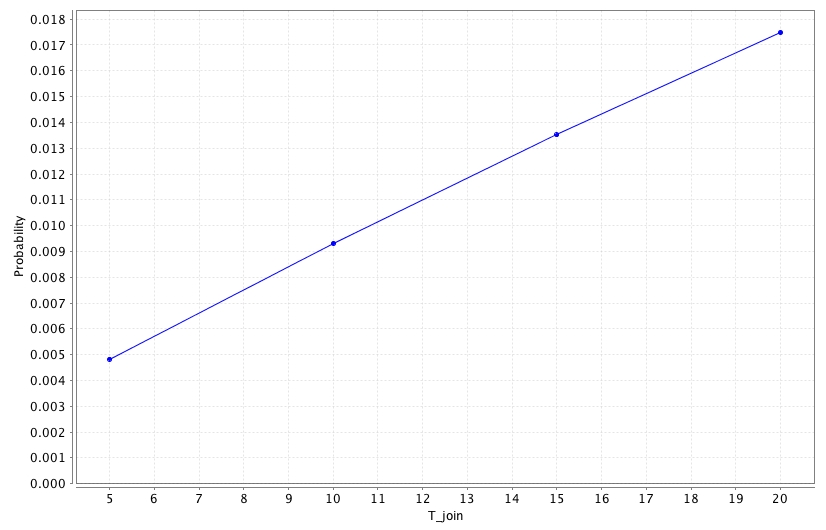} \\
      \small{c) Telecom Applications} & \small{d) Wireless Sensor Applications} \\
    \end{tabular}
  \label{tab:q2-j-2}
\end{figure}
\end{center}

Fig. \ref{tab:q2-t-1} and Fig. \ref{tab:q2-t-2} show the results for the \emph{time-based environment}, 
Fig. \ref{tab:q2-l-1} and Fig. \ref{tab:q2-l-2} show the results for the \emph{leave-based environment}, 
and Fig. \ref{tab:q2-j-1} and Fig. \ref{tab:q2-j-2} show the results for the \emph{join-based environment}.

The odd numbered figures contains the results for the scenarios HA, and SE; 
whereas the even numbered figures contains the results for the scenarios CBA, PHHC, TA, and WSA.

The x-axis always shows the key update threshold, that is {\tt T\_time}, {\tt T\_leave}, or {\tt T\_join} depending on the strategy. 
On the y-axis we have the steady-state probability that the key is compromised in the long run.  
In each graph, there is only one curve and it gives the steady-state probability for given key update thresholds.
Number of points in the curves are some times sparse due to time-consuming model checking operation when the state space is large, however all the graphics contain all the threshold values that were used in the corresponding graphics in the previous section, i.e. key confidentiality. As an example, for \emph{Home Automation} scenario and \emph{time-based strategy}, we had four different threshold values in the key confidentiality results at Fig. \ref{tab:q1-t-1}-a i.e. \{3,6,9,12\}. In this section, the relevant results are in Fig. \ref{tab:q2-t-1}-a where you can see all the threshold values from 1 to 12, obviously covers the values from key confidentiality.

In \textbf{time-based key update}, there is always a correlation between the threshold value and the steady-state probability. Obviously, an increase in the threshold value always results in an increase in the probability. All the scenarios show the same behaviour.

In \textbf{leave-based key update}, the situation is not much different than the time-based strategy in the sense that the correlation exist. However, one very slight difference is the change on the slope for the very first threshold values which is visible especially in CBA and TA scenarios as shown in Fig. \ref{tab:q2-l-2}-a and Fig. \ref{tab:q2-l-2}-c, respectively.

In \textbf{join-based key update}, completely the same as the leave-based strategy as graphical pattern. However, the numerical results are slightly higher than the results of the former strategy. still, the difference is so small that it is not clearly visible on the graphs. 
\clearpage
\pagebreak

\clearpage
\pagebreak

\clearpage
\pagebreak

\clearpage
\pagebreak
\subsubsection{Key Recovery}
\begin{center} 
\begin{figure}[!h]
  \caption{Results of question 3 for HA and SE in the time based environment}
    \begin{tabular}{cc}
      \includegraphics[width=8.5cm]{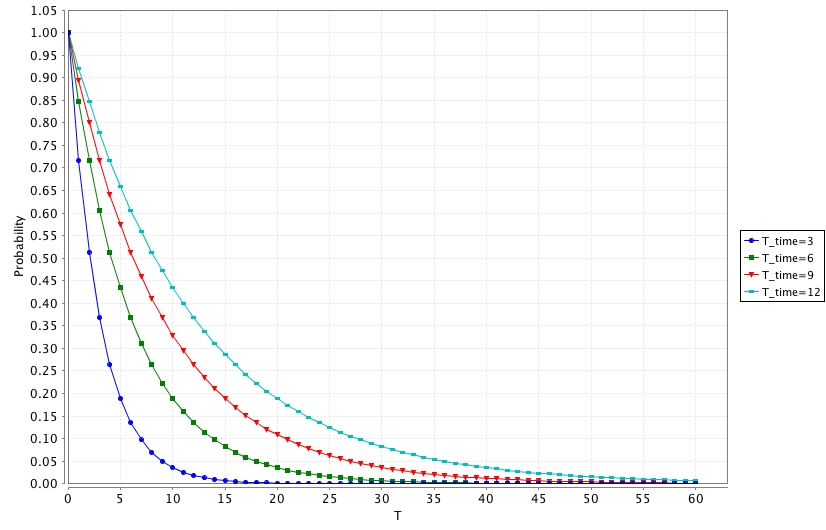} & \includegraphics[width=8.5cm]{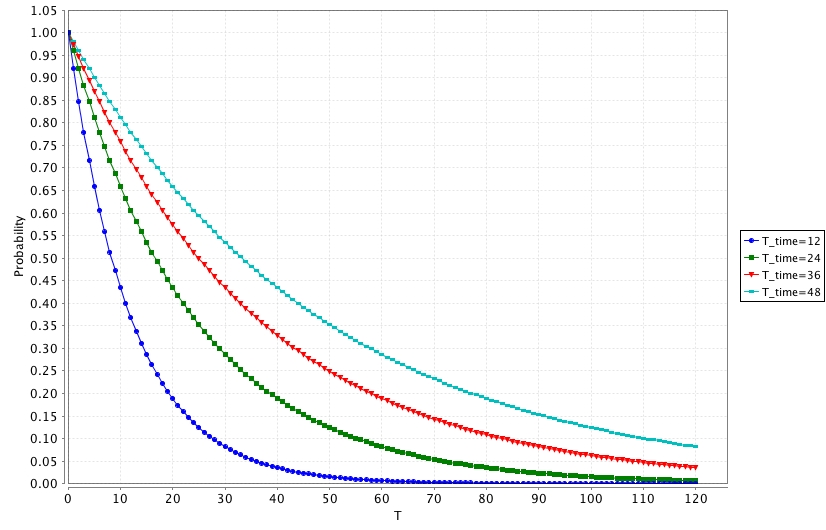}       \\
      \small{a) Home Automation} & \small{b) Smart Energy} \\
    \end{tabular}
  \label{tab:q3-t-1}
\end{figure}
\end{center}

\begin{center} 
\begin{figure}[!h]
  \caption{Results of question 3 for CBA, PHHC, TA, and WSA in the time based environment}
    \begin{tabular}{cc}
      \includegraphics[width=8.5cm]{q3-cba-tb} & \includegraphics[width=8.5cm]{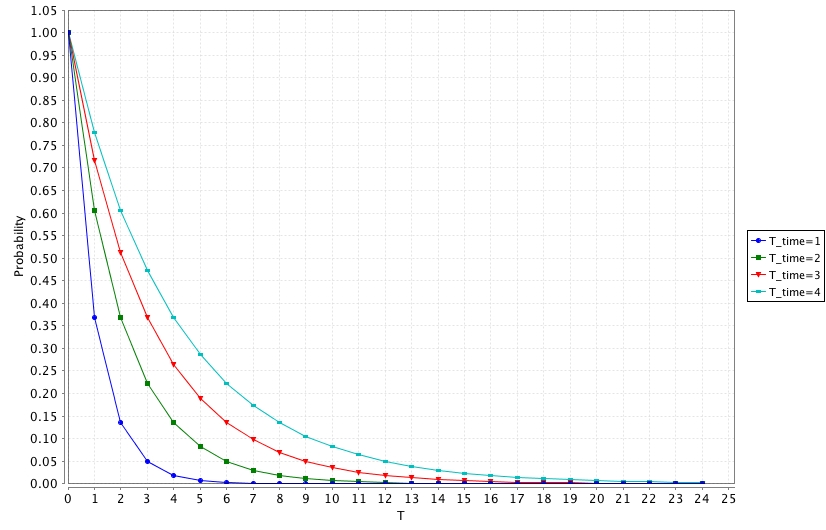}       \\
      \small{a) Commercial Building Automation} & \small{b) Personal, Home and Hospital Care} \\
      \includegraphics[width=8.5cm]{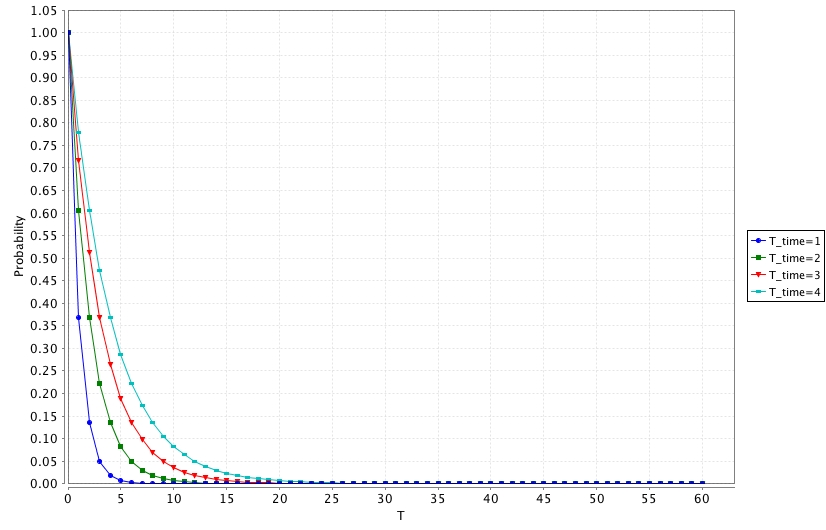} & \includegraphics[width=8.5cm]{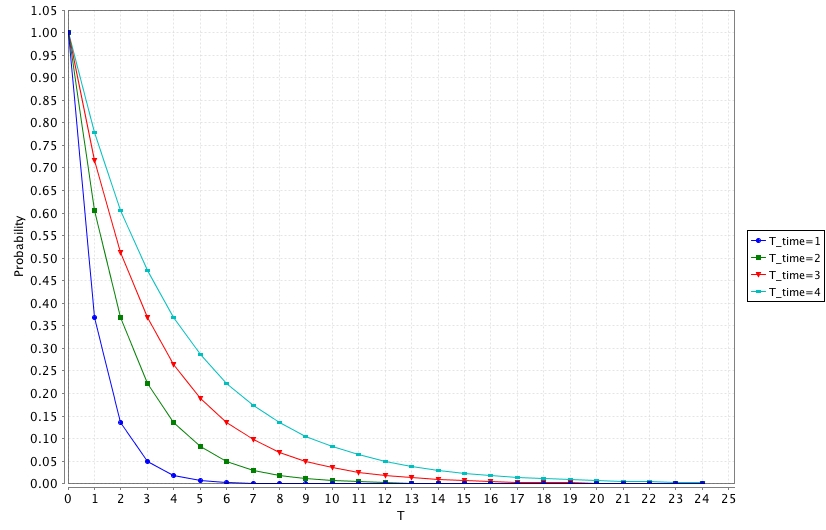}       \\
      \small{c) Telecom Applications} & \small{d) Wireless Sensor Applications} \\
    \end{tabular}
  \label{tab:q3-t-2}
\end{figure}
\end{center}

\begin{center} 
\begin{figure}[!h]
  \caption{Results of question 3 for HA and SE in the leave based environment}
    \begin{tabular}{cc}
      \includegraphics[width=8.5cm]{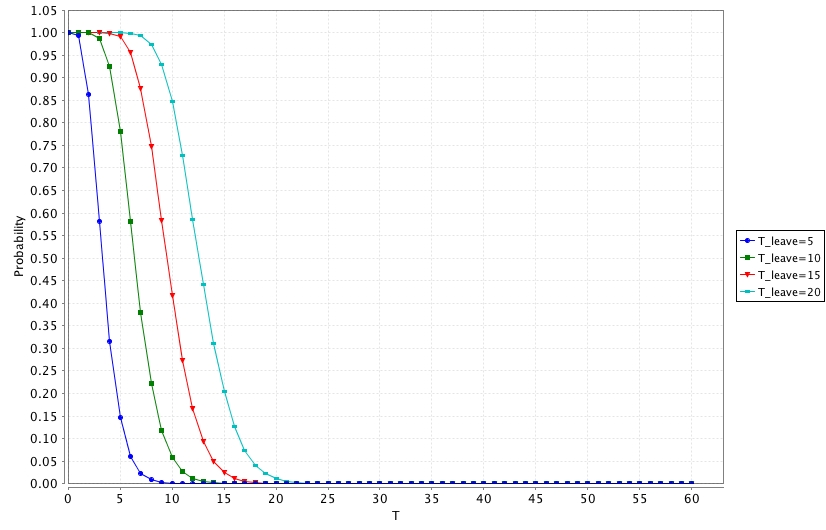} & \includegraphics[width=8.5cm]{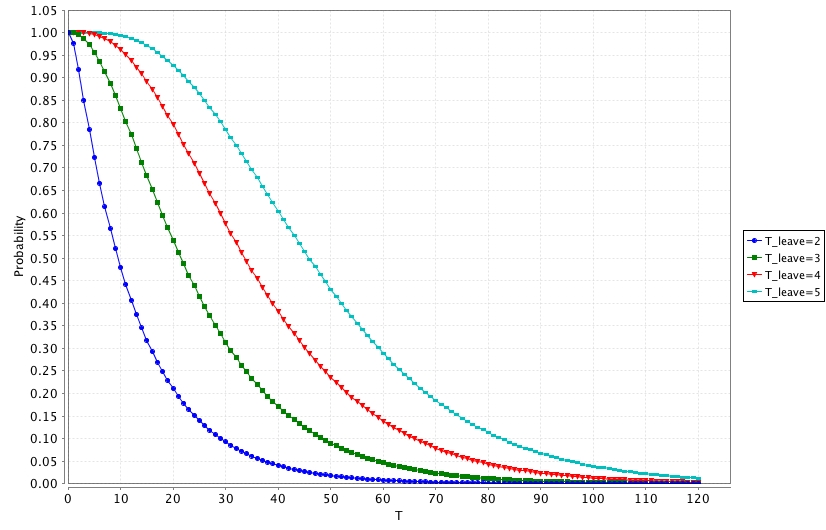}       \\
      \small{a) Home Automation} & \small{b) Smart Energy} \\
    \end{tabular}
  \label{tab:q3-l-1}
\end{figure}
\end{center}

\begin{center} 
\begin{figure}[!h]
  \caption{Results of question 3 for CBA, PHHC, TA, and WSA in the leave based environment}
    \begin{tabular}{cc}
      \includegraphics[width=8.5cm]{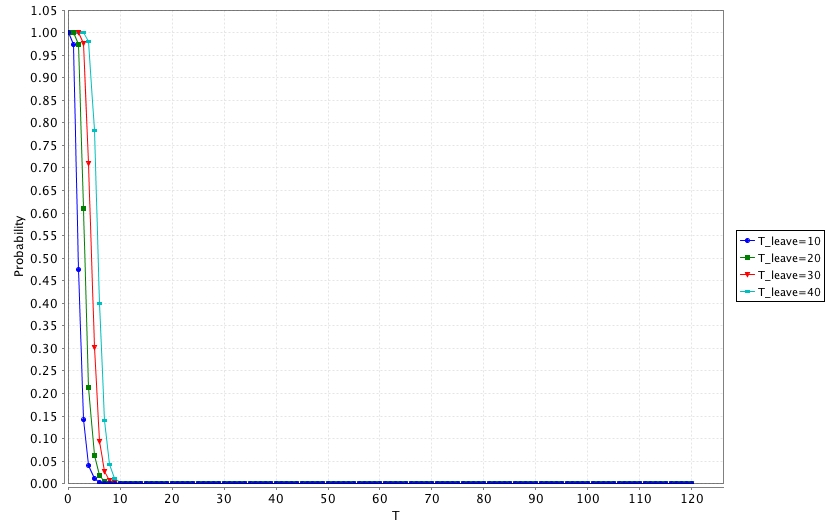} & \includegraphics[width=8.5cm]{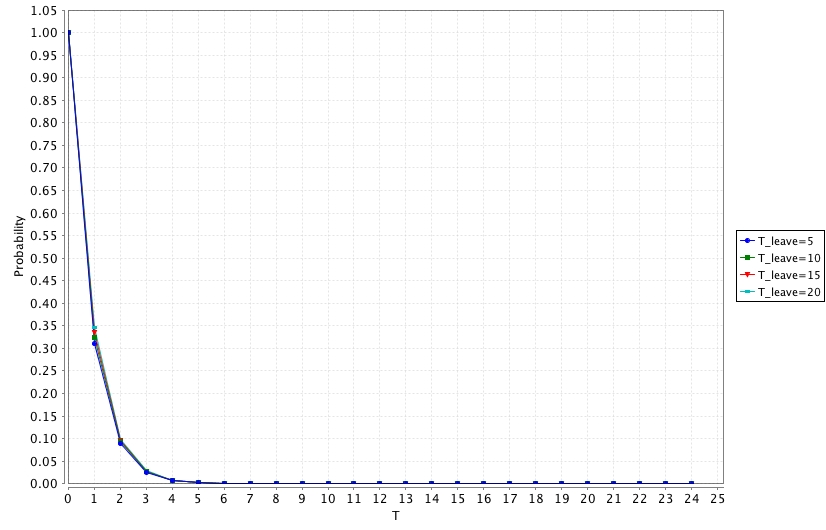}       \\
      \small{a) Commercial Building Automation} & \small{b) Personal, Home and Hospital Care} \\
      \includegraphics[width=8.5cm]{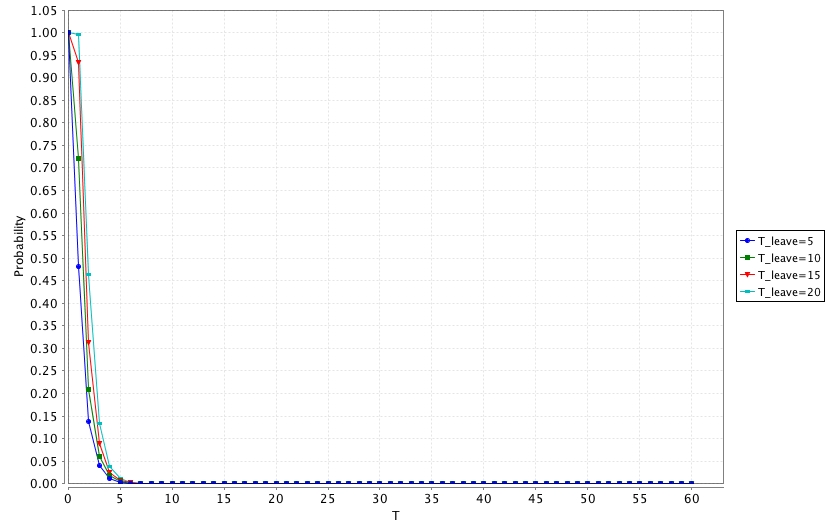} & \includegraphics[width=8.5cm]{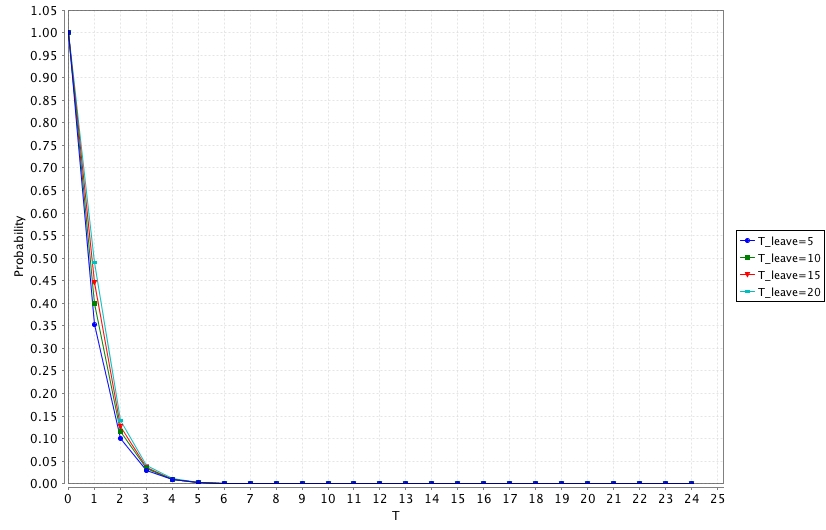}       \\
      \small{c) Telecom Applications} & \small{d) Wireless Sensor Applications} \\
    \end{tabular}
  \label{tab:q3-l-2}
\end{figure}
\end{center}

\begin{center} 
\begin{figure}[!h]
  \caption{Results of question 3 for HA and SE in the join based environment}
    \begin{tabular}{cc}
      \includegraphics[width=8.5cm]{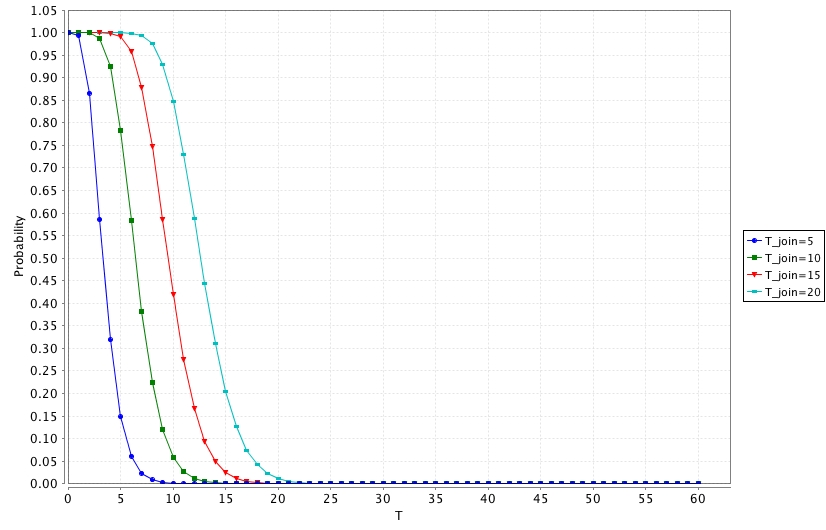} & \includegraphics[width=8.5cm]{q3-se-jb}       \\
      \small{a) Home Automation} & \small{b) Smart Energy} \\
    \end{tabular}
  \label{tab:q3-j-1}
\end{figure}
\end{center}

\begin{center} 
\begin{figure}[!h]
  \caption{Results of question 3 for CBA, PHHC, TA, and WSA in the join based environment}
    \begin{tabular}{cc}
      \includegraphics[width=8.5cm]{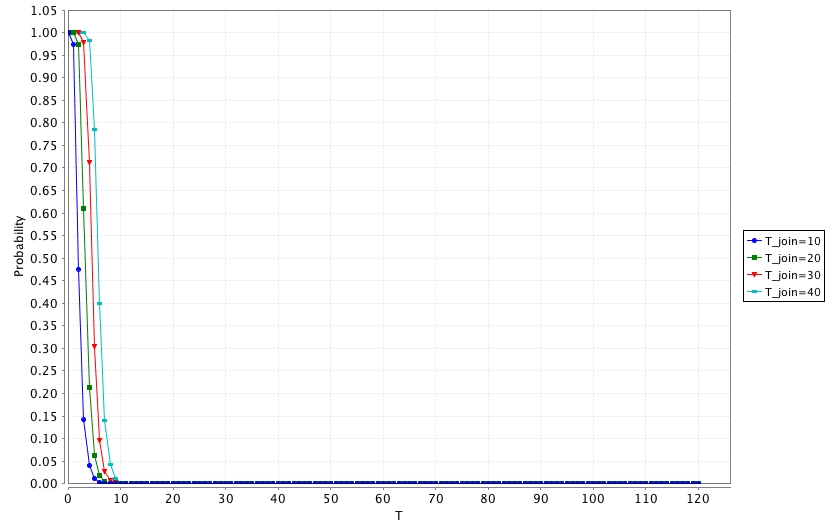} & \includegraphics[width=8.5cm]{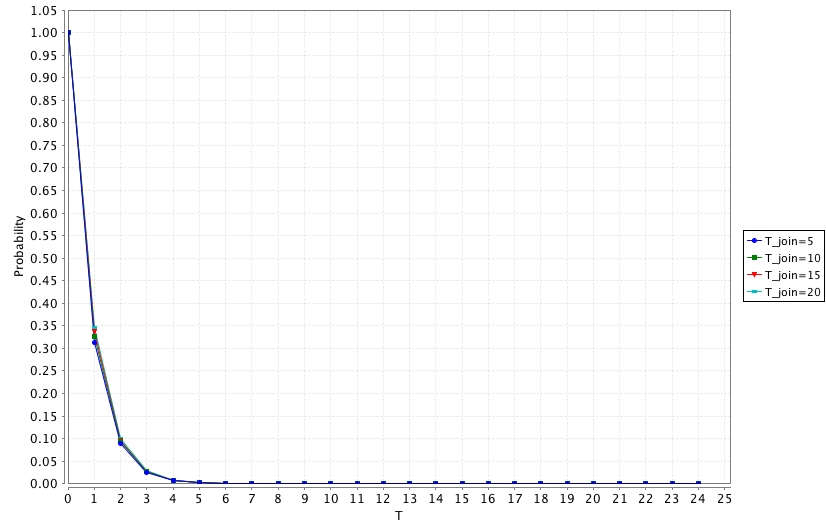} \\
      \small{a) Commercial Building Automation} & \small{b) Personal, Home and Hospital Care} \\
      \includegraphics[width=8.5cm]{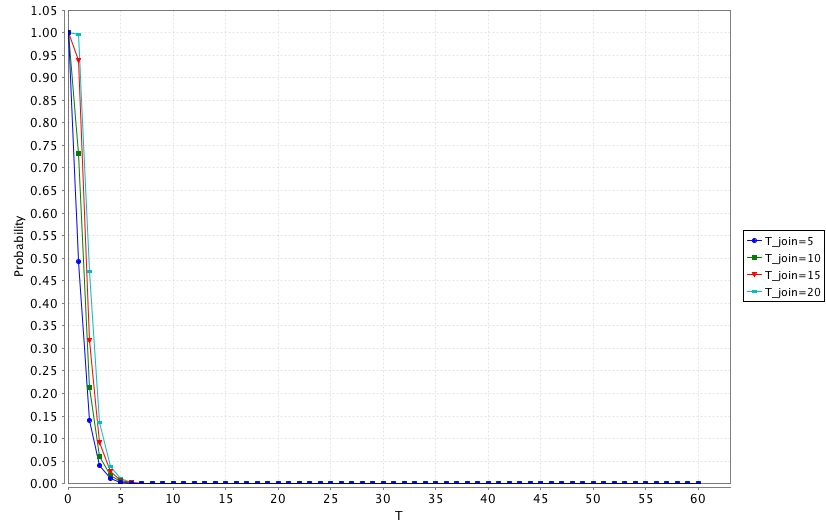} & \includegraphics[width=8.5cm]{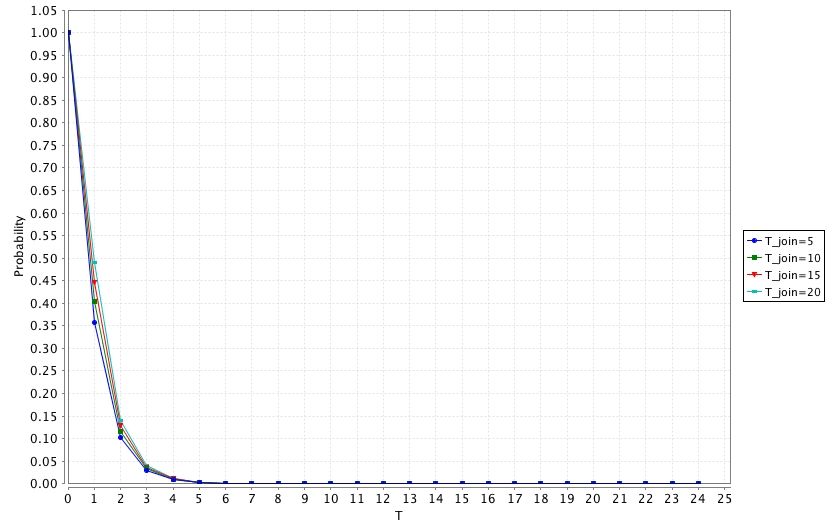} \\
      \small{c) Telecom Applications} & \small{d) Wireless Sensor Applications} \\
    \end{tabular}
  \label{tab:q3-j-2}
\end{figure}
\end{center}

Fig. \ref{tab:q3-t-1} and Fig. \ref{tab:q3-t-2} show the results for the \emph{time-based environment}, 
Fig. \ref{tab:q3-l-1} and Fig. \ref{tab:q3-l-2} show the results for the \emph{leave-based environment}, 
and Fig. \ref{tab:q3-j-1} and Fig. \ref{tab:q3-j-2} show the results for the \emph{join-based environment}.

The odd numbered figures contains the results for the scenarios HA, and SE; 
whereas the even numbered figures contains the results for the scenarios CBA, PHHC, TA, and WSA.

The x-axis always shows the time, specifically month {\tt T} of interest. 
On the y-axis we have the probability (or risk) that it takes more than {\tt T} months to recover from a compromised key situation.
The curves represent different intervals {\tt T\_time} for resetting the key, depending on the selected scenario. 
For example, a curve labeled as {\tt T\_leave=10} means that the key is updated after 10 devices leave (or 10 device-leave actions) following the last key update. 

In \textbf{time-based key update}, the probability starts at a maximum point, and then starts to decline with a very steep slope. the probability declines until it is negligible, however the speed of drop slows down as {\tt T} grows. This behaviour of the systems, is independent of the selection of the scenario nor the key update threshold value. 
We can briefly say that key compromise recovery probability becomes 0 eventually, and how long it takes for it depends on 1) the key update scenario, and 2) the threshold value.

In \textbf{leave-based key update}, the behaviour of the system is slightly different than the time-based key update. Even though the probability starts with maximum and then goes to minimum as it is in the former strategy, in the beginning the slop is not steep for a while. This property is clearly visible in scenarios such as HA and SE (see in Fig. \ref{tab:q3-l-1}-a and Fig. \ref{tab:q3-l-1}-b), and almost invisible in scenarios such as PHHC and WSA (see in Fig. \ref{tab:q3-l-2}-b and Fig. \ref{tab:q3-l-2}-d). Other than that, the duration of the initial high probability phase changes depending on the threshold value.

In \textbf{join-based key update}, is pretty much similar to the leave-based key update in terms of the patterns in graphical results. However, the numerical probabilities are always slightly different than the leave-based key update. 
\clearpage
\pagebreak

\clearpage
\pagebreak

\clearpage
\pagebreak

\clearpage
\pagebreak
\subsubsection{Efficiency}
\begin{center} 
\begin{figure}[!h]
  \caption{Results of question 4 for HA and SE in the time based environment}
    \begin{tabular}{cc}
      \includegraphics[width=8.5cm]{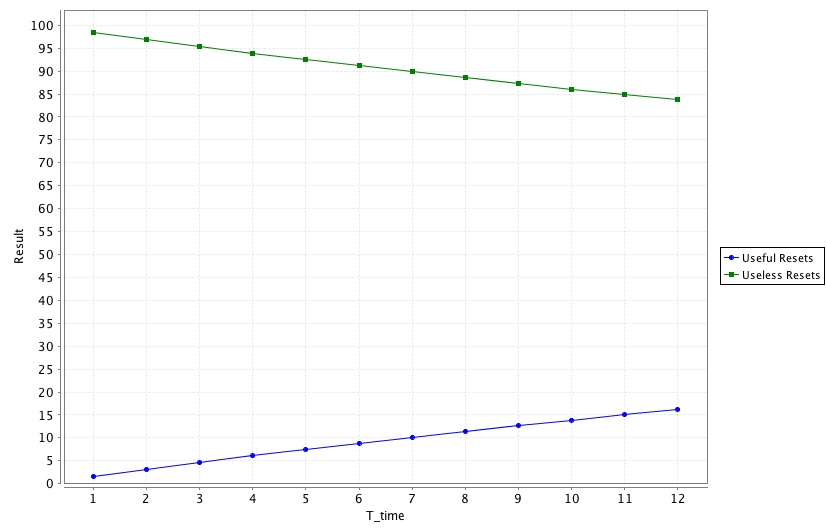} & \includegraphics[width=8.5cm]{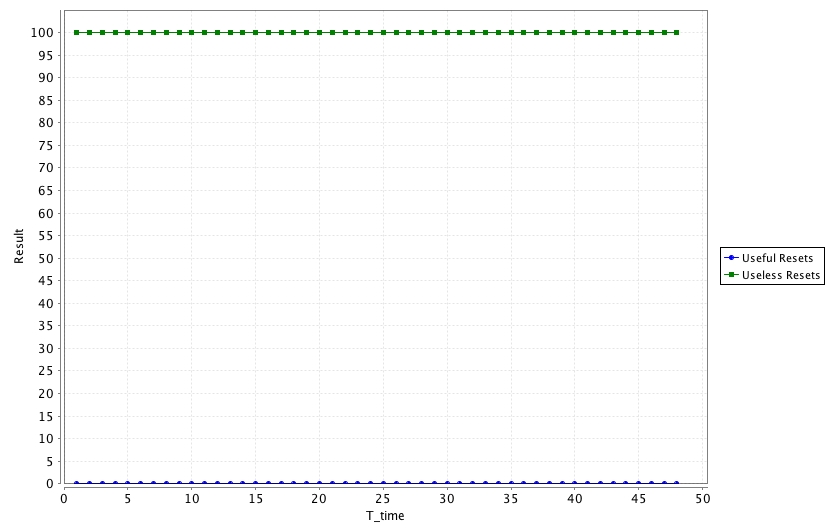}       \\
      \small{a) Home Automation} & \small{b) Smart Energy} \\
    \end{tabular}
  \label{tab:q4-t-1}
\end{figure}
\end{center}

\begin{center} 
\begin{figure}[!h]
  \caption{Results of question 4 for CBA, PHHC, TA, and WSA in the time based environment}
    \begin{tabular}{cc}
      \includegraphics[width=8.5cm]{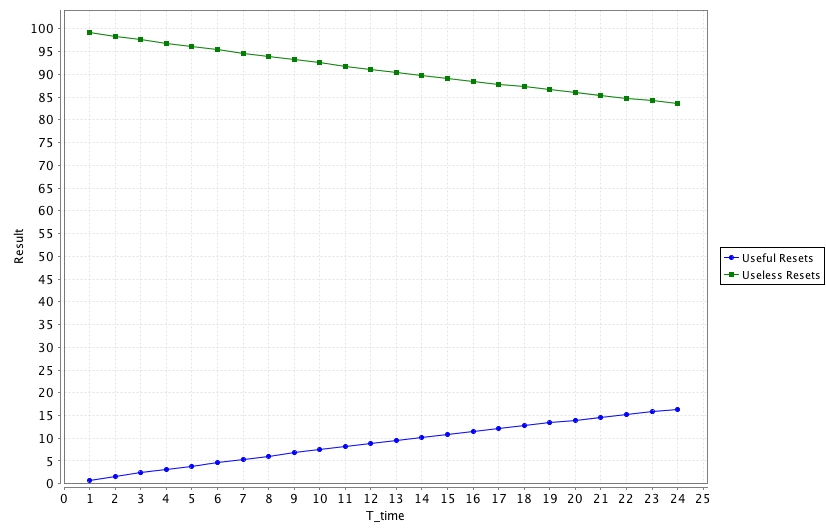} & \includegraphics[width=8.5cm]{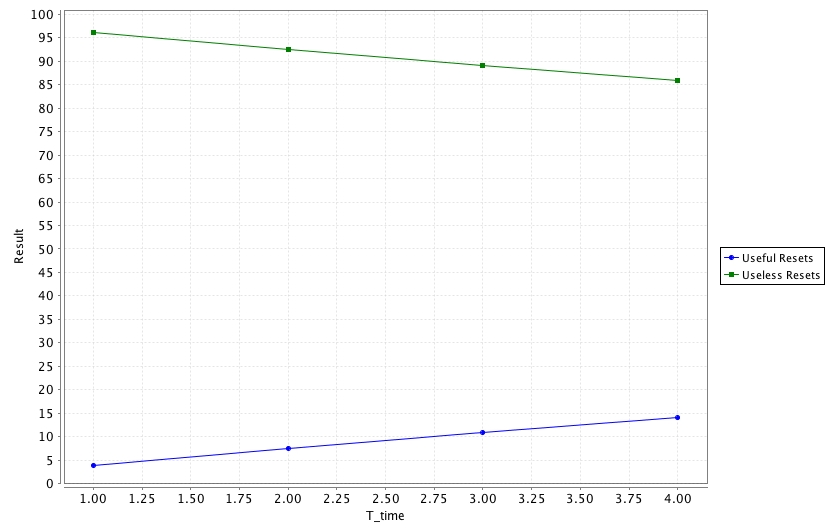}       \\
      \small{a) Commercial Building Automation} & \small{b) Personal, Home and Hospital Care} \\
      \includegraphics[width=8.5cm]{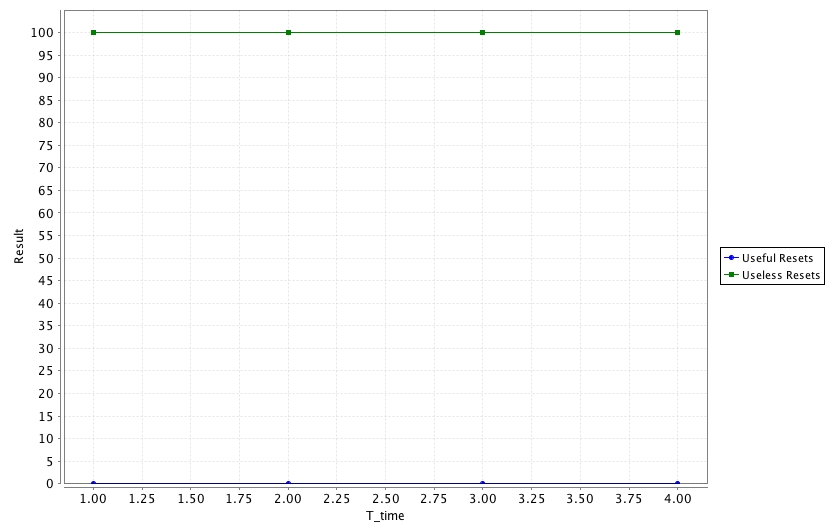} & \includegraphics[width=8.5cm]{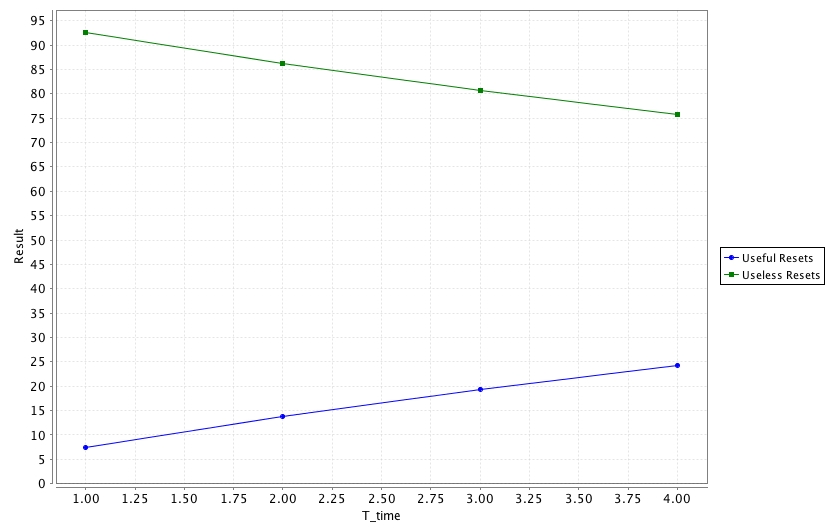}       \\
      \small{c) Telecom Applications} & \small{d) Wireless Sensor Applications} \\
    \end{tabular}
  \label{tab:q4-t-2}
\end{figure}
\end{center}

\begin{center} 
\begin{figure}[!h]
  \caption{Results of question 4 for HA and SE in the leave based environment}
    \begin{tabular}{cc}
      \includegraphics[width=8.5cm]{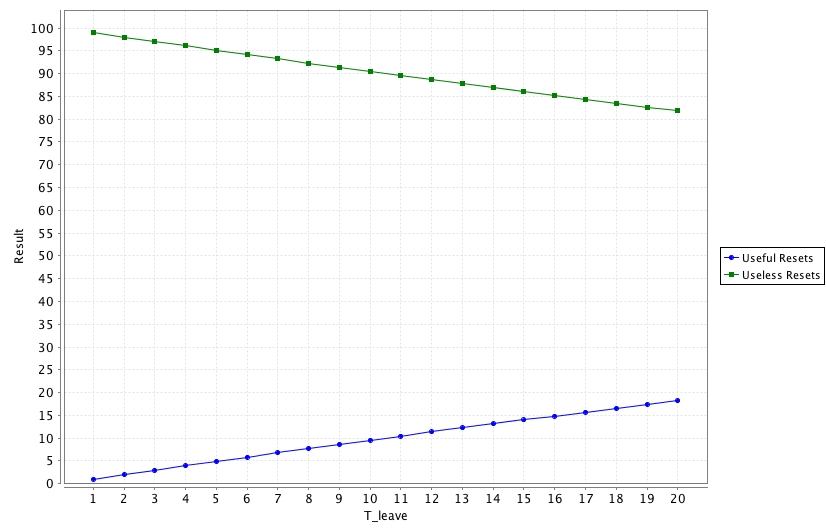} & \includegraphics[width=8.5cm]{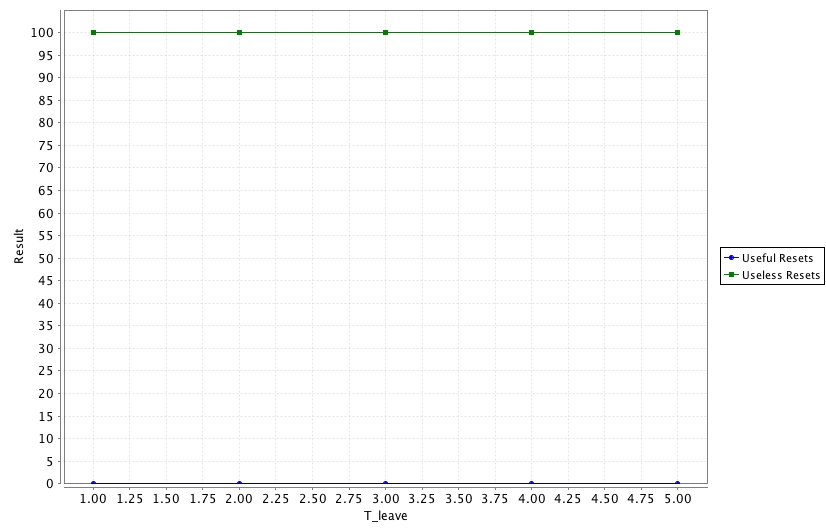}       \\
      \small{a) Home Automation} & \small{b) Smart Energy} \\
    \end{tabular}
  \label{tab:q4-l-1}
\end{figure}
\end{center}

\begin{center} 
\begin{figure}[!h]
  \caption{Results of question 4 for CBA, PHHC, TA, and WSA in the leave based environment}
    \begin{tabular}{cc}
      \includegraphics[width=8.5cm]{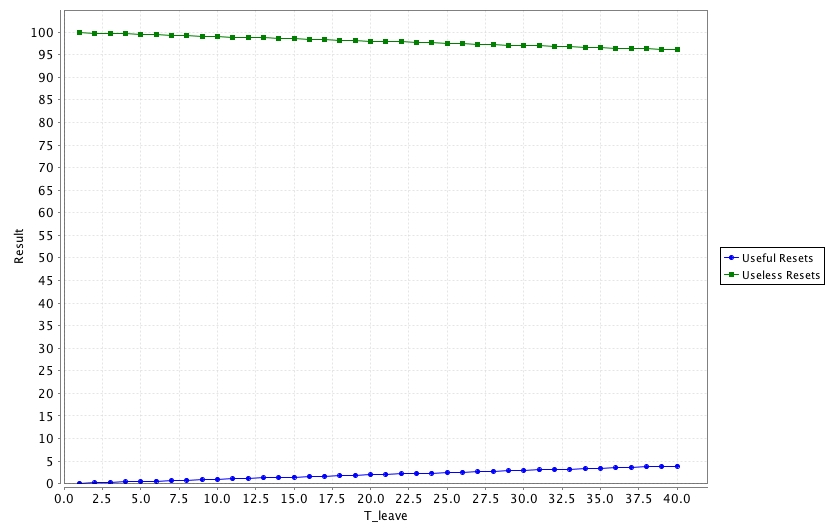} & \includegraphics[width=8.5cm]{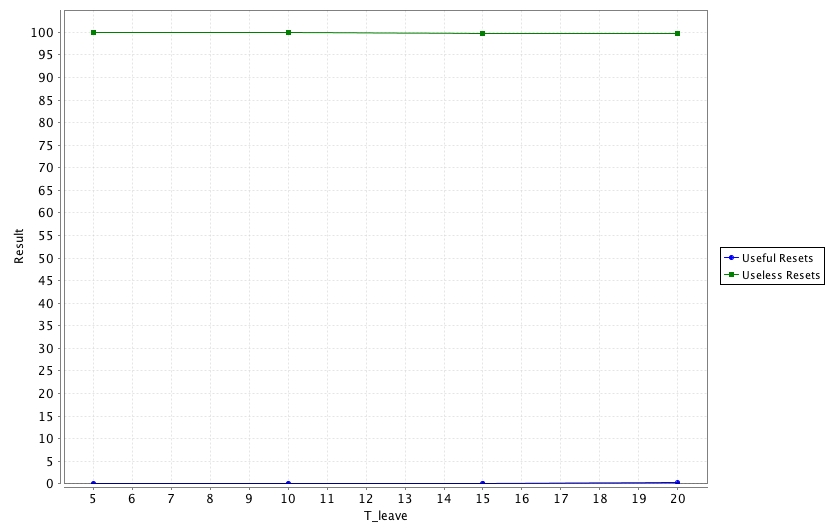}       \\
      \small{a) Commercial Building Automation} & \small{b) Personal, Home and Hospital Care} \\
      \includegraphics[width=8.5cm]{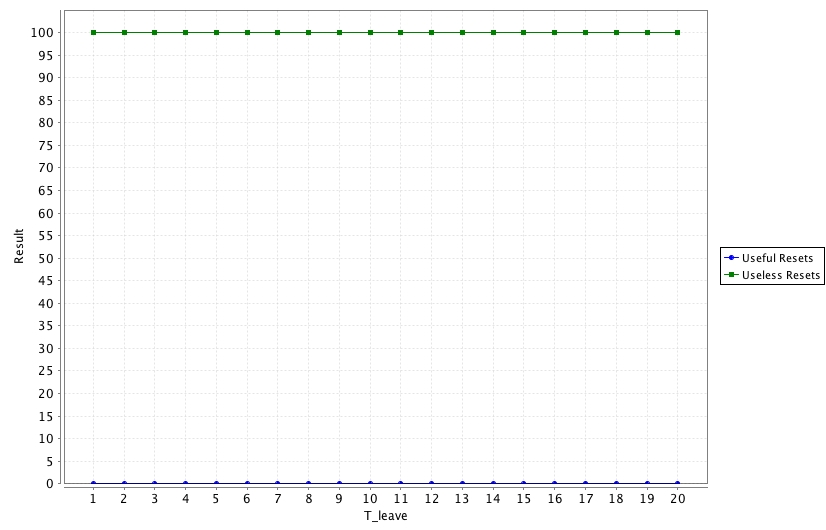} & \includegraphics[width=8.5cm]{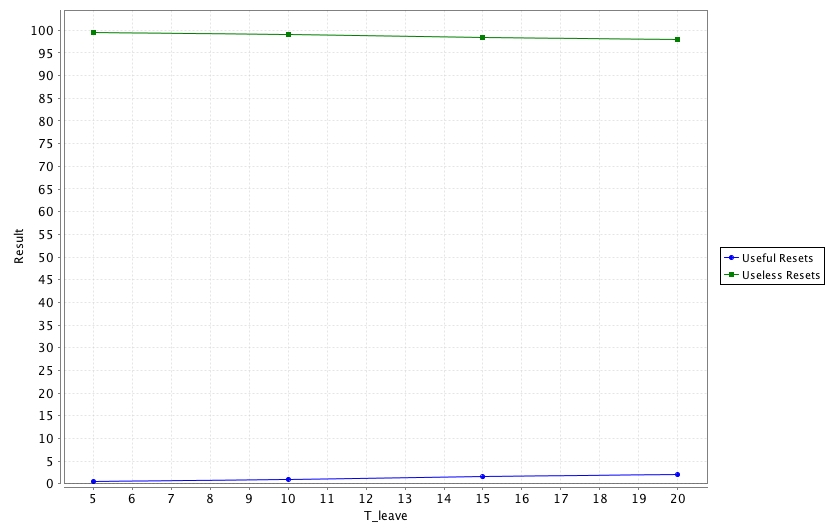}       \\
      \small{c) Telecom Applications} & \small{d) Wireless Sensor Applications} \\
    \end{tabular}
  \label{tab:q4-l-2}
\end{figure}
\end{center}

\begin{center} 
\begin{figure}[!h]
  \caption{Results of question 4 for HA and SE in the join based environment}
    \begin{tabular}{cc}
      \includegraphics[width=8.5cm]{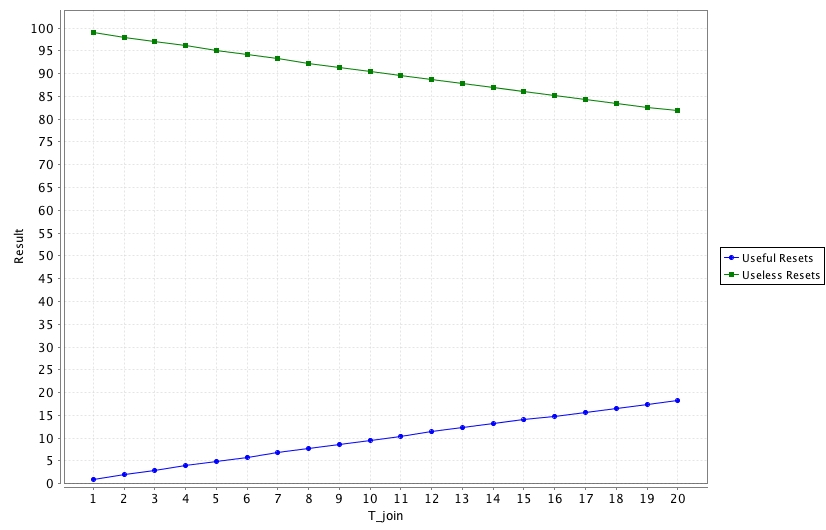} & \includegraphics[width=8.5cm]{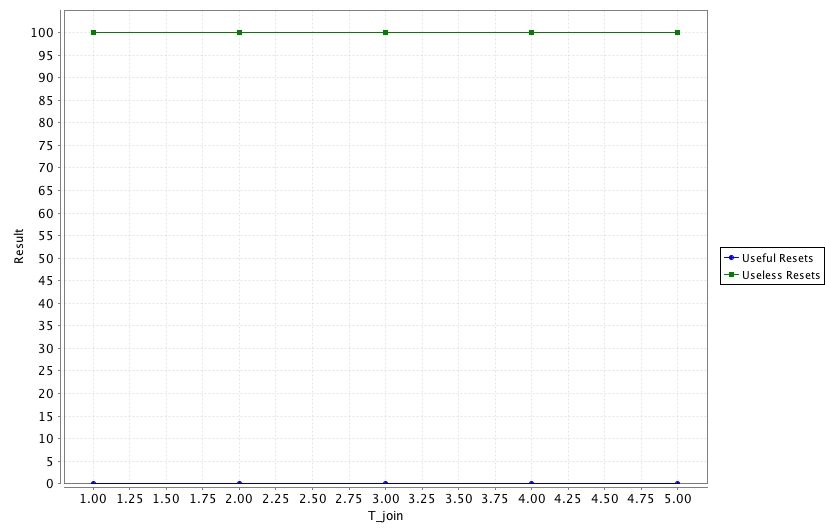}       \\
      \small{a) Home Automation} & \small{b) Smart Energy} \\
    \end{tabular}
  \label{tab:q4-j-1}
\end{figure}
\end{center}

\begin{center} 
\begin{figure}[!h]
  \caption{Results of question 4 for CBA, PHHC, TA, and WSA in the join based environment}
    \begin{tabular}{cc}
      \includegraphics[width=8.5cm]{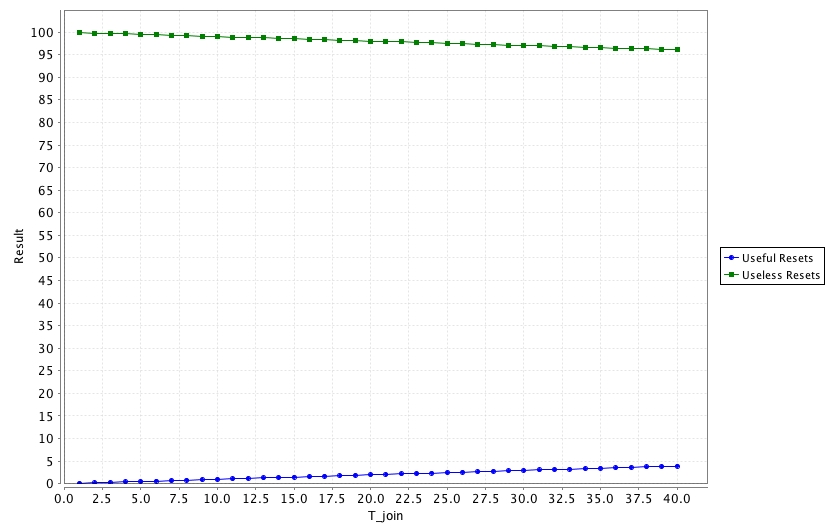} & \includegraphics[width=8.5cm]{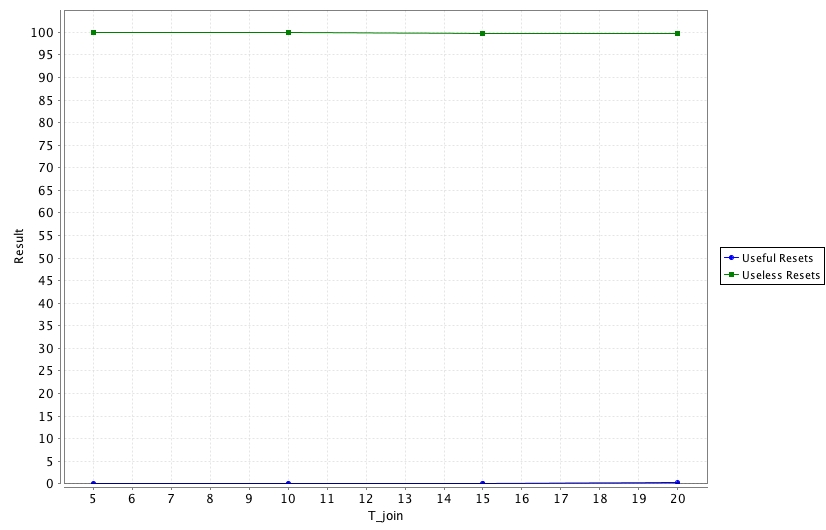} \\
      \small{a) Commercial Building Automation} & \small{b) Personal, Home and Hospital Care} \\
      \includegraphics[width=8.5cm]{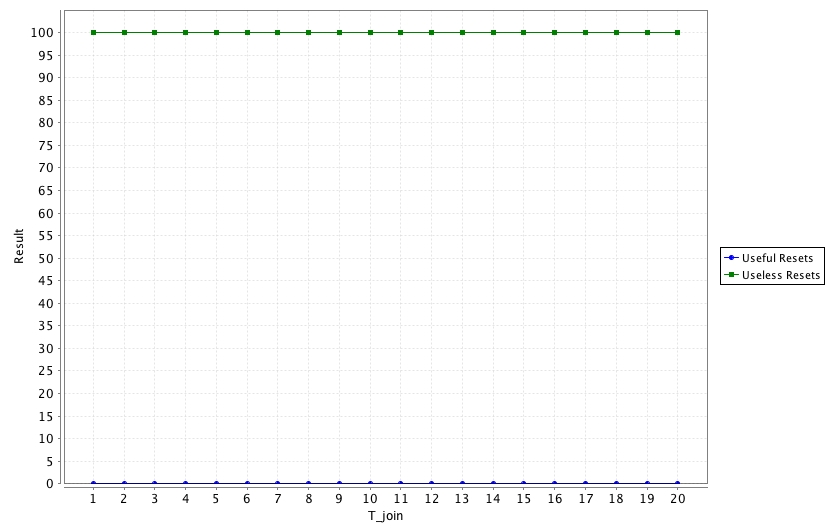} & \includegraphics[width=8.5cm]{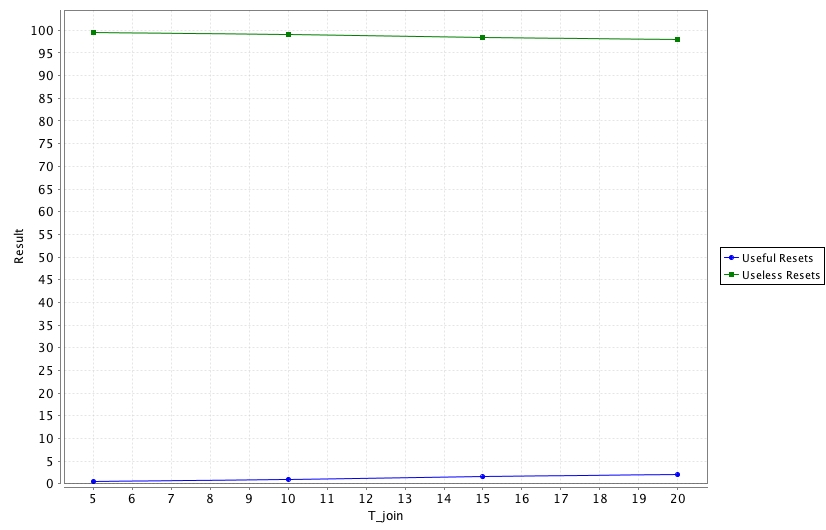} \\
      \small{c) Telecom Applications} & \small{d) Wireless Sensor Applications} \\
    \end{tabular}
  \label{tab:q4-j-2}
\end{figure}
\end{center}

Fig. \ref{tab:q4-t-1} and Fig. \ref{tab:q4-t-2} show the results for the \emph{time-based environment}, 
Fig. \ref{tab:q4-l-1} and Fig. \ref{tab:q4-l-2} show the results for the \emph{leave-based environment}, 
and Fig. \ref{tab:q4-j-1} and Fig. \ref{tab:q4-j-2} show the results for the \emph{join-based environment}.

The odd numbered figures contains the results for the scenarios HA, and SE; 
whereas the even numbered figures contains the results for the scenarios CBA, PHHC, TA, and WSA.

The x-axis always shows the key update threshold, that is {\tt T\_time}, {\tt T\_leave}, or {\tt T\_join} depending on the strategy. 
On the y-axis we have the percentages of the key updates in the long run.  
We classify a key update as a \emph{useful} key update if it recovers a key compromise situation, and \emph{useless} otherwise. 
In each graph, there are exactly two curves where one of them corresponds to the useful key updates, and the other to the useless ones.
Number of points in the curves depend on the number of thresholds we are trying.

In \textbf{time-based key update}, it is always the case that the useless key updates start at its maximum and useful ones start at their minimum percentage. Besides, percentage of useless key  updates always drop and useful ones always increase. When we try sufficiently large interval of threshold values we found out that these two curves (actually lines) always intersect and after some threshold point it is always the case that the percentage of the useful key updates get larger than the useless ones. However, picking the threshold value so large that it will make more useful key updates is not a rational choice since it actually means sacrificing security for performance (actually power consumption). Therefore, in our experiments you can see that in most of the cases increase and decrease in percentages are very little, and in some cases (such as SE and TA application profiles) the change is hardly visible.

In \textbf{leave-based key update}, we got almost the same behaviour with the time-based key update except the changes on the numerical results (percentages), off course. Besides, since the threshold sets in these two scenarios does not give exactly same confidentiality and long run probability results; such differences in the values are no surprise.

In \textbf{join-based key update}, we figure out that all the results are as we got in the leave-based key update. It is hard to make a distinction and define a winner between leave-based and join-based strategies just by looking at the results for key update efficiency. 
\clearpage
\pagebreak

\clearpage
\pagebreak

\clearpage
\pagebreak

